\documentclass[10pt, twocolumn, a4paper, longbibliography]{revtex4-1}

\usepackage{amsmath,amssymb}
\usepackage[a4paper, top=2.0cm, bottom=1.5cm, left=1.5cm, right=1.5cm]{geometry}
\usepackage{helvet}
\usepackage{graphicx}
\usepackage{color}
\usepackage[colorlinks=true,bookmarks=false,citecolor=blue,urlcolor=blue]{hyperref}
\usepackage{physics}

\newcommand{\eps}{\varepsilon}

\begin{document}

\preprint{APS/123-QED}

\title{Higher-order topological states \\ mediated by long-range coupling in $D_4$-symmetric lattices}

\author{Nikita A. Olekhno}
\affiliation{Department of Physics and Engineering, ITMO University, Saint Petersburg 197101, Russia}
\author{Alina D. Rozenblit}
\affiliation{Department of Physics and Engineering, ITMO University, Saint Petersburg 197101, Russia}
\author{Valerii I. Kachin}
\affiliation{Department of Physics and Engineering, ITMO University, Saint Petersburg 197101, Russia}
\author{Alexey A. Dmitriev}
\affiliation{Department of Physics and Engineering, ITMO University, Saint Petersburg 197101, Russia}
\author{Oleg I. Burmistrov}
\affiliation{Department of Physics and Engineering, ITMO University, Saint Petersburg 197101, Russia}
\author{Pavel S. Seregin}
\affiliation{Department of Physics and Engineering, ITMO University, Saint Petersburg 197101, Russia}
\author{Dmitry V. Zhirihin}
\affiliation{Department of Physics and Engineering, ITMO University, Saint Petersburg 197101, Russia}
\author{Maxim A. Gorlach}
\affiliation{Department of Physics and Engineering, ITMO University, Saint Petersburg 197101, Russia}


\maketitle

\small

\textbf{Topological physics opens a door towards flexible routing and resilient localization of waves of various nature. Recently proposed higher-order topological insulators~\cite{2017_Benalcazar_Science,2018_Schindler} provide advanced control over wave localization in the structures of different dimensionality. In many cases, the formation of such higher-order topological phases is governed by the lattice symmetries, with kagome~\cite{2018_Xue,2018_Ni_NatureMat} and breathing honeycomb~\cite{2015_Wu} lattices being prominent examples. Here, we design and experimentally realize the resonant electric circuit with $D_4$ symmetry and additional next-nearest-neighbor couplings. As we prove, a coupling of the distant neighbors gives rise to an in-gap corner state. Retrieving the associated invariant directly from the experiment, we demonstrate the topological nature of the designed system, revealing the role of long-range interactions in the formation of topological phases. Our results thus highlight the distinctions between tight-binding systems and their photonic counterparts with long-range couplings.
}

Higher-order topological insulators have recently emerged as a distinct class of topological systems implemented experimentally with various platforms, including  crystalline solids~\cite{2018_Schindler_Bismuth}, phononic~\cite{2018_Serra_Garcia}, acoustic~\cite{2018_Xue,2018_Ni_NatureMat}, and electromagnetic setups working at infrared~\cite{2019_Mittal,2019_Hassan} and microwave~\cite{2018_Peterson,2020_Li} frequencies, as well as resonant electric circuits~\cite{2018_Imhof,2019_Serra_Garcia}. Due to their ability to confine field in the structures of different dimensionality, such higher-order topological phases are promising candidates for topological resonators and lasers~\cite{2017_Bahari, 2020_Zhang, 2020_Han, 2020_Kim}.

In many cases, the physics of such systems can be understood in terms of tight-binding models involving only the nearest neighbors' interaction. However, this is not the case for photonics, where the long-range interactions of the individual meta-atoms can significantly alter the band structure~\cite{2020_Li}.

\begin{figure}[b]
    \centerline{\includegraphics[width=8cm]{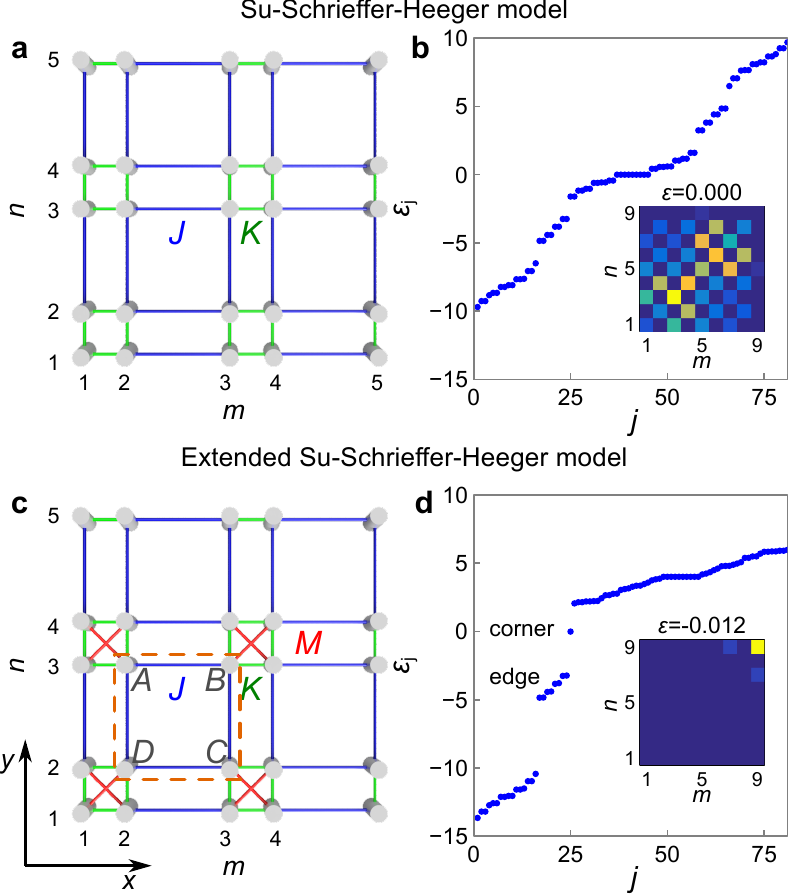}}
    \caption{\footnotesize {\bf Extended two-dimensional Su-Schrieffer-Heeger model}. {\bf a}: Schematics of the two-dimensional SSH model realized as an array of nearest-neighbor-coupled resonators with the coupling strengths $J>0$ and $K>J$. {\bf b}: Spectrum of energies $\varepsilon_{j}$ versus eigenvalue number $j$ for the model $9\times 9$ sites from panel {\bf a} with couplings $J=1$, $K=4$. Inset shows the wavefunction for the eigenmode with $\eps=0$. {\bf c}: Proposed extension of 2D SSH with additional couplings $M>0$ in the strong-link unit cell. Orange dashed line shows the weak-link unit cell choice used for the analysis of a periodic system. Labels $A$, $B$, $C$, and $D$ denote four sites of the unit cell. {\bf d}: Energy spectrum of the model in panel {\bf c} with parameters $J=1$, $K=M=4$ having the size of $9 \times 9$ sites. Inset shows the field profile of the corner mode.}
    \label{fig:Model}
\end{figure}

Recently, several microwave experiments~\cite{2019_Chen,2019_Xie} have demonstrated the emergence of corner states in the two-dimensional generalization of the celebrated Su-Schrieffer-Heeger model (SSH) with $D_4$ symmetry~\cite{2017_Wakabayashi}. At the same time, the respective tight-binding model [Fig.~\ref{fig:Model}a] does not feature a zero-energy bandgap, and the associated corner state coexists with the continuum of the bulk modes [Fig.~\ref{fig:Model}b].

In this Letter, we prove that the formation of zero-energy bandgap hosting topological corner states in $D_4$-symmetric systems crucially depends on the next-nearest-neighbor interaction, which facilitates the emergence of higher-order topological phase. To isolate the physics related to the next-nearest-neighbor coupling, we design and fabricate a sample based on a resonant LC circuit, where the magnitude of the coupling parameters can be flexibly controlled [Fig.~\ref{fig:Model}c]. Besides the retrieval of frequencies and mode profiles of bulk, edge, and corner states, we also reveal generalized chiral symmetry of the model and calculate the topological invariant associated with $D_4$ lattice symmetry.


\begin{figure*}[t]
    \centerline{\includegraphics[width=17cm]{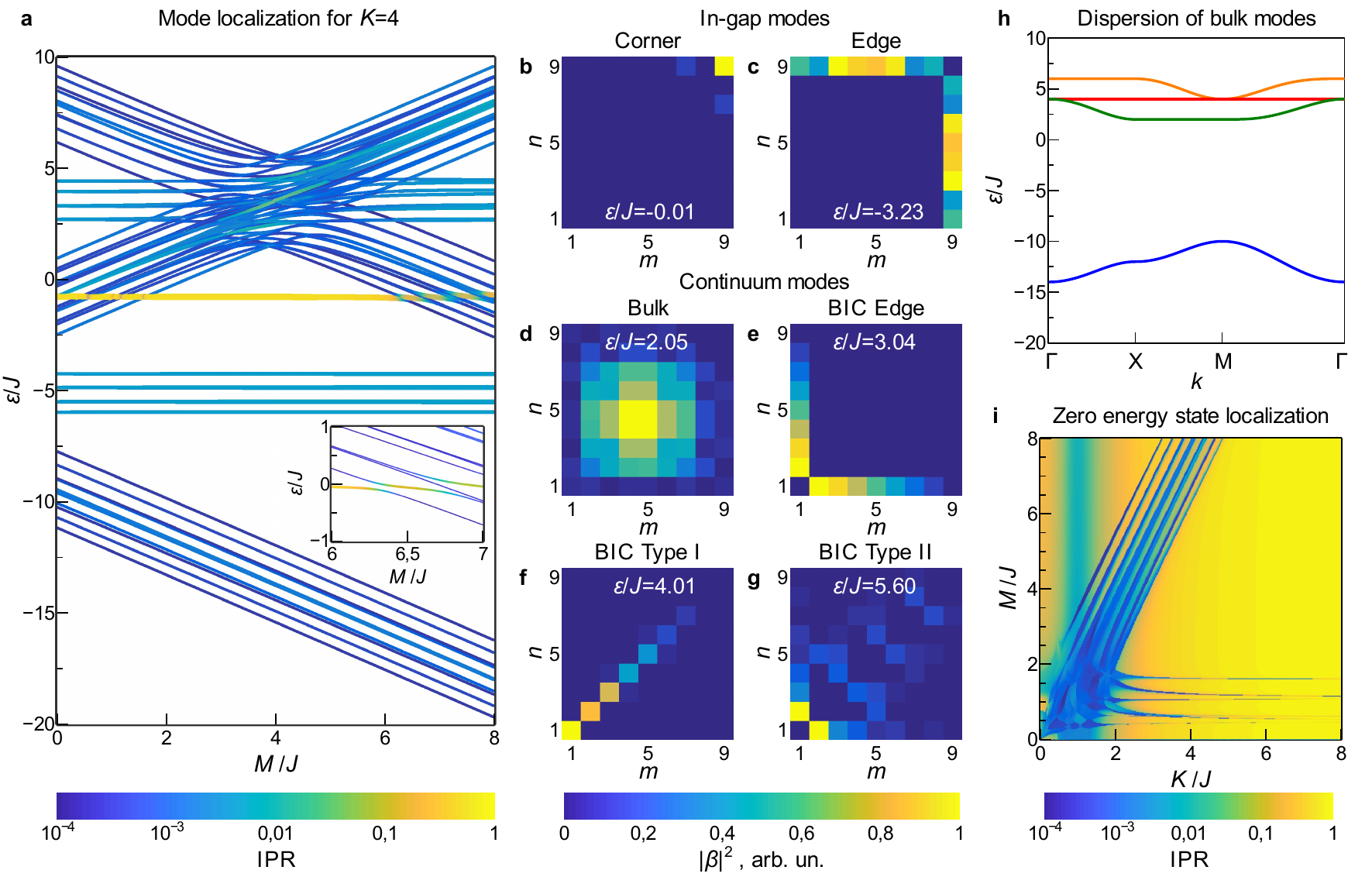}}
    \caption{\footnotesize {\bf Eigenmodes of the proposed $D_4$-symmetric model.} {\bf a}: Energy spectrum of  $9 \times 9$ structure with coupling constants $J=1$, $K=4$ versus next-nearest-neighbor coupling $M$ in the range $0<M<8$. Color shows the inverse participation ratio (IPR) of the eigenmodes defined by Eq.~\eqref{eq:IPR}. Inset demonstrates the enlarged fragment of spectrum showing avoided crossings and the formation of hybrid modes upon the interaction of topological corner state with continuum modes. {\bf b-c}: Eigenmode profiles corresponding to in-gap corner state ({\bf b}) and edge state ({\bf c}) at $M=4$. {\bf d}: delocalized bulk state at $M=4$. {\bf e-g}: Eigenmode profiles for bound states in the continuum (BIC). {\bf e}: edge state in the continuum at $M=3.999$. {\bf f}: BIC Type I corner state at $M=4.007$. {\bf g}: BIC Type II corner state at $M=5.085$. {\bf h}: Dispersion diagram for bulk bands of the periodic lattice with couplings $J=1$, $K=M=4$ for the wave vector changing along the $\Gamma-X-M-\Gamma$ trajectory. Different colors correspond to four bulk modes. {\bf i}: Colorplot for the inverse participation ratio calculated for the state with the energy closest to zero for $9 \times 9$ system with $J=1$ as a function of couplings $K$ and $M$.}
    \label{fig:Eigenmodes}
\end{figure*}

The eigenstates of both described models [Fig.~\ref{fig:Model}a,c] are found as the solutions to the eigenvalue problem
\begin{equation}
    \sum_{m',n'}H_{mn,m'n'}\beta_{m'n'} = \varepsilon\,\beta_{mn}\:,
    \label{eq:Tight_binding}
\end{equation}
where $\beta_{mn}$ coefficients describe the amplitude of the field at $(m,n)$ site, $\eps$ is the mode energy defined such that the zero energy corresponds to the resonance frequency of an isolated site, while the Hamiltonian matrix $\hat{H}$ embeds the properties of the system. The nonzero elements of the Hamiltonian $-J$, $-K$, or $-M$ correspond to the coupling links between the respective sites $(m,n)$ and $(m',n')$ as further discussed in the Methods section. Without loss of generality, we set smaller coupling constant $J=1$, whereas $K>J$.

Solving the eigenvalue problem Eq.~\eqref{eq:Tight_binding}, we recover the spectra of both systems, without and with next-nearest-neighbor coupling $M$, depicted in Figs.~\ref{fig:Model}b,d, respectively. Regardless of the ratio $K/J$, the canonical two-dimensional (2D) SSH is gapless near the zero energy, and thus the corner state coexists with the continuum of bulk modes [Fig.~\ref{fig:Model}b]. However, diagonal couplings $M$ within each strongly coupled unit cell open a bandgap and yield a spectrally isolated corner-localized state.
It should be stressed that the proposed system [Fig.~\ref{fig:Model}c] is the minimal model which captures the effect of long-range interactions in photonic systems since the diagonal links $M$ introduced in the strong coupling unit cell are the dominant terms related to the next-nearest-neighbor interaction. Even though the corner state profile shown in the inset of Fig.~\ref{fig:Model}d strongly resembles that in the canonical quadrupole insulator~\cite{2017_Benalcazar_Science}, all coupling links here are positive, which vastly simplifies the experimental implementation of the proposed system.


To quantify the localization properties of the eigenmodes in our model, we evaluate their inverse participation ratios (IPR) \cite{1974_Thouless, 2020_Mukherjee}
\begin{equation}
    {\rm IPR} = \sum_{n,m}|\beta_{mn}|^4,
    \label{eq:IPR}
\end{equation}
where the summation is performed over all sites $(m,n)$ of the lattice $1\leq m,n\leq N$, and the eigenmode profile is normalized by the condition $\sum_{n,m} |\beta_{mn}|^2=1$. There are three scenarios of IPR scaling with the increase of the system size $N$. If the mode is spread over the entire system, the superposition coefficients $\beta_{mn}\propto 1/N$ and hence $IPR\propto 1/N^2$. If the eigenstate is confined to the system edge, then $\beta_{mn}\propto 1/\sqrt{N}$, and $IPR\propto 1/N$. Finally, if the mode is localized at the corner, only few $\beta_{mn}$ contribute to the wave function, hence $IPR\approx 1$.

\begin{figure*}[t]
    \centerline{\includegraphics[width=17cm]{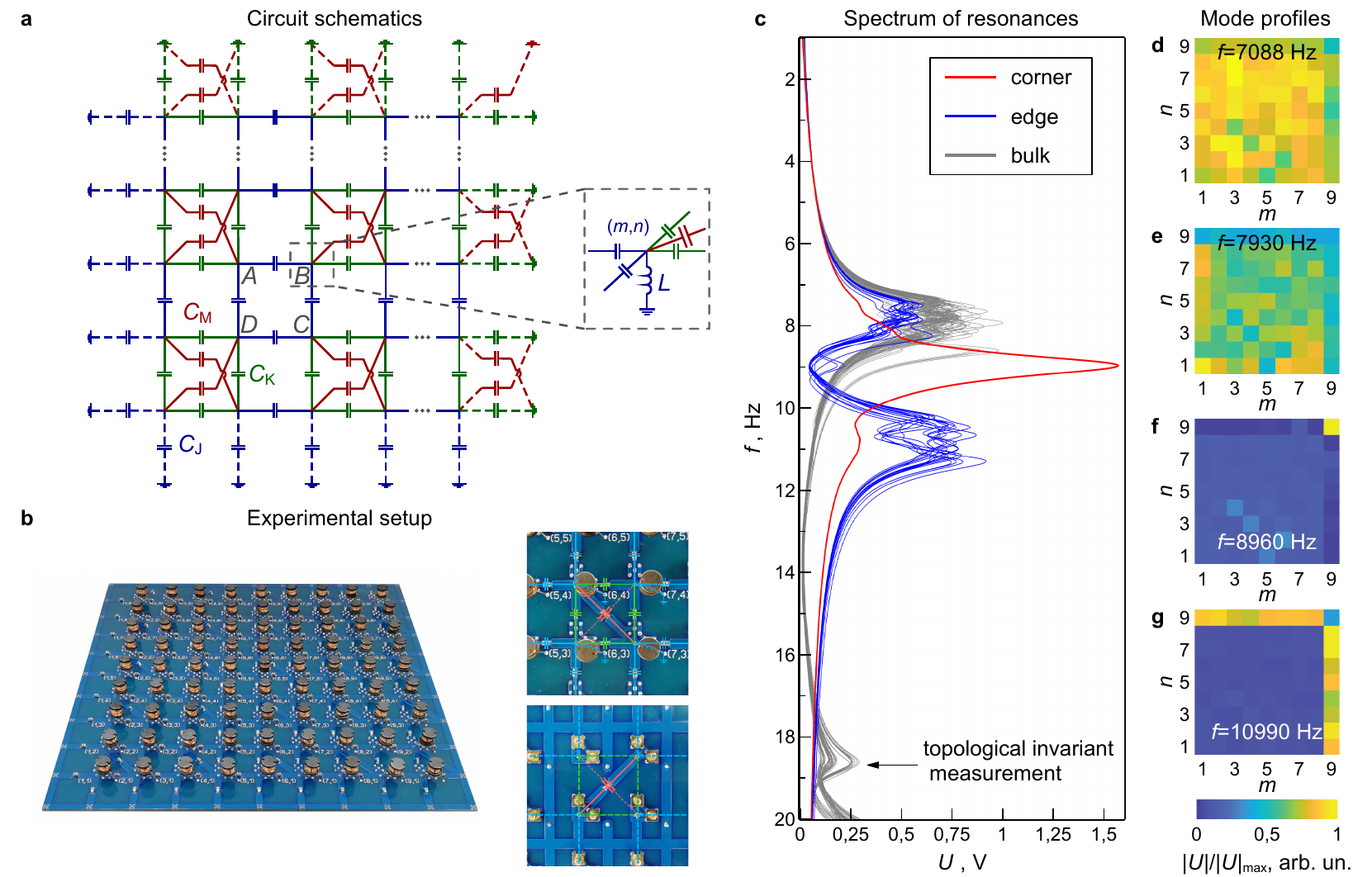}}
    \caption{\footnotesize {\bf Electric circuit realization.} {\bf a}: Equivalent electric circuit realizing the tight-binding model Fig.~\ref{fig:Model}c. Each node is grounded by the inductor $L$ and connected to its neighbors via capacitors $C_{\rm J}$, $C_{\rm K}$ and $C_{\rm M}$ representing respective tunneling links $J$, $K$ and $M$ of the extended 2D SSH model. The boundary nodes of the circuit are grounded with additional elements to provide the exact mapping between the Kirchhoff's rules for the circuit and tight-binding equations. Labels $A$, $B$, $C$, and $D$ denote sites of the unit cell in accordance with Fig.~\ref{fig:Model}c. {\bf b}: The photograph of the experimental $9 \times 9$ circuit. Half of the additional diagonal couplings $C_{\rm M}$ are seen at the top side of PCB along with capacitors $C_{\rm J}$, $C_{\rm K}$ and grounding inductors $L$, while the remaining couplings $C_{\rm M}$ are placed at the bottom side carrying also the plugs to connect the measurement equipment. The unit cell of the circuit is shown in the insets to the right. The upper inset demonstrates top view, while the bottom inset shows the opposite side of PCB mirrored to show the matching of the bonds. {\bf c}: Resonant response of the circuit measured at every node  in the range from $1$ to $20$ kHz showing the on-site voltages $U(f)$ between given node $(m,n)$ and ground excited by the external harmonic driving at frequency $f$ with the amplitude $U_{\rm ext}=63$~mV applied between the corresponding node and ground. Grey curves represent bulk and edge nodes with coordinates $1 \le m,n \le 8$, blue curves correspond to the edge nodes with $1 \le m \le 8$, $n=9$ and $1 \le n \le 8$, $m=9$, red curve represents the corner node $m=9$, $n=9$. The arrow points towards the frequency $f=18616$~Hz at which the topological invariant is retrieved. {\bf d-g} Spatial maps of the on-site voltage responses to the external excitation $U_{\rm ext}=63$~mV at a given frequency $f$ representing bulk ({\bf d},{\bf e}), corner ({\bf f}), and edge ({\bf g}) states of the extended SSH model. Color shows the absolute value of voltage between node $(m,n)$ and ground normalized by its maximal value calculated for each map separately.}
    \label{fig:Circuit}
\end{figure*}

Thus, increasing the system size, we expect to observe three distinct types of participation ratio scaling. This intuition is confirmed by Fig.~\ref{fig:Eigenmodes}a, which shows the evolution of the spectrum with the increase of the next-nearest neighbor coupling. Three distinct colors present in the diagram are directly associated with the three types of localization: bright yellow corresponds to the corner state, teal blue color shows the edge states, whereas dark blue depicts bulk states.

The results in Fig.~\ref{fig:Eigenmodes}a suggest that the corner state is spectrally isolated only for the certain range of next-nearest-neighbor coupling strengths $M_{\rm min}<M<M_{\rm max}$, with $M_{\rm min} \approx 1.6$ and $M_{\rm max} \approx 6.3$ for $K=4$. The corner state profile in such a case is depicted in Fig.~\ref{fig:Eigenmodes}b featuring a pronounced localization at the corner with the weak coupling links $J$.

The emergence of the corner state in our system is accompanied by the formation of the edge states [Fig.~\ref{fig:Eigenmodes}c] inherited from the 2D Su-Schrieffer-Heeger model and pinned to the edges terminated by the weak links. Note that the edge states' energy remains unaffected by the next-nearest-neighbor coupling $M$ as long as the edge states remain confined to the edges where the next-nearest-neighbor coupling is absent [Fig.~\ref{fig:Eigenmodes}a].

At the same time, the energies of the bulk modes delocalized over the entire 2D system [Fig.~\ref{fig:Eigenmodes}d] feature a pronounced dependence on $M$. As a result, the bands of bulk and edge states can cross for some parameters interacting with each other and giving rise to more exotic localization types including bound states in the continuum (BIC)~\cite{2016_Hsu}. Interestingly, such BIC states arising in the avoided crossing region of bulk and edge modes can localize at the strong link edges [Fig.~\ref{fig:Eigenmodes}e] or even at the strong link corner [Fig.~\ref{fig:Eigenmodes}f,g] in agreement with the prediction of symmetry-protected BIC in the conventional 2D SSH~\cite{2020_Cerjan}. Specifically, the strong link corner hosts two states with different behavior under reflection relative to $m=n$ line: symmetric [Fig.~\ref{fig:Eigenmodes}f] and antisymmetric [Fig.~\ref{fig:Eigenmodes}g]. We refer to them as type I and type II BIC corner states, respectively, in analogy to the recent work on photonic kagome lattice~\cite{2020_Li}.

It should be stressed that the BIC Type II corner state appears less localized than BIC Type I. Therefore, for a small $9 \times 9$ system considered here, it can be misinterpreted as a bulk excitation. However, analysis of a larger system allows us to prove the localized nature of the mode (Supplementary Note 4).

To probe the topological properties of our model, we examine the bulk bands of a periodic system with a four-site unit cell giving rise to the four bulk bands. While the bulk modes' dispersion can be derived analytically (Supplementary Note 1), the energies and the field profiles of these modes satisfy {\it generalized chiral symmetry} resembling that in kagome lattice~\cite{2018_Ni_NatureMat}. In particular, a sum of eigenvalues corresponding to the four bulk bands of our system is equal to zero [Fig.~\ref{fig:Eigenmodes}h], and the respective mode profiles are linked to each other via the generalized chiral symmetry operator (see Methods).

Having the energies and the field profiles of the bulk modes, we now assess the topological characteristics of our model by checking the behavior of the field profiles under $C_2$ or $C_4$ symmetry transformations in few high-symmetry points of the first Brillouin zone~\cite{2019_Benalcazar}. Due to the $C_4$ symmetry of the lattice, the topological invariant contains three independent components $\chi = (\#X_1^{(2)}-\#\Gamma_1^{(2)}, \#M_1^{(4)}-\#\Gamma_1^{(4)}, \#M_2^{(4)}-\#\Gamma_2^{(4)})$, where the upper index denotes the type of the applied rotation operator ($C_2$ or $C_4$), lower index describes the behavior of the wave function under the symmetry transformation and $\#$ denotes the number of eigenstates with a given transformation law below the particular bandgap in $\Gamma$, $M$ or $X$ point of the first Brillouin zone.

Similar to the SSH model case, the topological invariant depends on the choice of the unit cell. If the unit cell is chosen with the strong links inside, the topological invariant is $(0,0,0)$, indicating the absence of topological states at the strong link corner. However, if the unit cell is chosen with weak links inside [Fig.~\ref{fig:Model}c], the topological invariant appears to be nonzero
\begin{equation}
    \chi = (-1, -1, 0),
\end{equation}
heralding the emergence of higher-order topological corner state with associated corner charge $Q_{\rm corner}=\frac{1}{4}$ and dipole polarization ${\bf P}=(\frac{1}{2},\frac{1}{2})$ \cite{2019_Benalcazar}. It should be stressed that the topological invariant does not depend on $M$. Nevertheless, next-nearest-neighbor interaction is crucial to open the bandgap at energies close to zero.

Once the topological origin of the corner state is confirmed, we focus on its localization properties. To this end, we trace the evolution of the inverse participation ratio (IPR) of the fixed corner mode in the $9\times 9$ system when the dimerization strength $K/J$ and the next-nearest-neighbor coupling $M/J$ are varied. The calculated phase diagram is shown in Fig.~\ref{fig:Eigenmodes}i.


We observe that even weak additional couplings $M$ readily yield localized states for certain values of the strong coupling constant $K$. However, the localization deteriorates significantly once the corner mode falls into the continuum of bulk states. Despite the small size of the array, the phase diagram features quite a complicated structure, thus highlighting the rich physics of the proposed model.

To experimentally confirm that the next-nearest-neighbor couplings $M$ provide the crucial ingredient in the formation of in-gap topological corner state, we need to eliminate the contribution of other spurious long-range couplings that inevitably arise in optical or microwave setups based on resonator arrays. To this end, we construct a topological electric circuit, Fig.~\ref{fig:Circuit}a,b, in which we can directly control the couplings between the sites by placing the desired lumped elements. This extreme flexibility in managing the geometry and amplitudes of the couplings in comparison with the other platforms allows applying electric circuits to emulate such exotic phenomena as four-dimensional quantum Hall phase \cite{2020_Wang}, two-particle topological states of interacting photons \cite{2020_Olekhno}, and nonlinearity-induced topological states \cite{2018_Hadad} along with the implementation of higher-order topological insulators \cite{2018_Imhof, 2019_Bao, 2020_Liu} and edge states in topological insulators \cite{2015_Ningyuan}, including the standard two-dimensional SSH model \cite{2019_Liu}.

The construction of electric circuit model is based upon the exact correspondence between the initial tight-binding problem Eq.(\ref{eq:Tight_binding}) describing on-site amplitudes $\beta_{mn}$ and a set of Kirchhoff's rules describing electric potentials $\varphi_{mn}$ at the respective sites of the equivalent circuit depicted in Fig.~\ref{fig:Circuit}a. The link between the parameters of the circuit such as capacitances $C_{\rm J}$, $C_{\rm K}$, $C_{\rm M}$, and grounding inductors $L$ from one side and the parameters in the tight-binding model from the other reads:
\begin{equation}
    K=\frac{C_{\rm K}}{C_{\rm J}},~M=\frac{C_{\rm M}}{C_{\rm J}},~\varepsilon=\frac{f_{0}^2}{f^2}-\left(2+2\frac{C_{K}}{C_{J}}+\frac{C_{M}}{C_{J}}\right)\:,
    \label{eq:Circuit_Parameters}
\end{equation}
where $f$ is the frequency of the circuit mode, $\eps$ is the energy in the tight-binding model and $f_{0}=1/(2\pi\,\sqrt{LC_{\rm J}})$  (see Methods for details). Thus, ascending tight-binding energies $\eps$ correspond to the descending mode frequencies $f$ of the electric circuit, which is exploited further.


The experimental realization of the circuit with $C_{\rm J}=1\,\mu{\rm F}$, $C_{\rm K}=C_{\rm M}=4\,\mu{\rm F}$, and $L = 23.2\,\mu{\rm H}$ corresponding to the considered model with $K=M=4$ and the size of $9 \times 9$ sites is shown in Fig.~\ref{fig:Circuit}b. Such a circuit has resonances in the kHz frequency range. To probe the modes of the circuit, we apply the external harmonic signal at frequency $f$ with amplitude $U_{\rm ext}=63~{\rm mV}$, attaching the signal generator having series impedance $50~{\rm Ohm}$ to the given node and ground. Then, we measure the resulting voltage between this node and ground, which characterizes the circuit impedance.

The spectroscopy of the circuit shown in Fig.~\ref{fig:Circuit}c reveals a bandgap between $9$ and $18$~kHz occupied by the modes in the range $10-12$ kHz localized at the edges of the circuit, and a single mode pinned to the site $(9,9)$ with the frequency around $9$~kHz. Attaching harmonic signal generator to every node of the circuit and measuring the voltages between the given node $(m,n)$ and the ground at a fixed frequency $f$, we recover voltage maps shown in Fig.~\ref{fig:Circuit}d-g. As seen from these maps, the respective modes represent bulk, edge, and corner states in the considered extended SSH model. The obtained positions of the resonant peaks agree with the results expected from the tight-binding model.

The peaks in the spectrum experience considerable broadening caused by ohmic losses in the inductors and wires of the printed circuit board. Another reason for broadening is the spread in lumped elements' values, as discussed in Supplementary Note 6. It should be stressed that the in-gap corner state Fig.~\ref{fig:Circuit}f possesses the largest Q-factor compared to the other resonances in the circuit, reaching $Q \approx 10$. It also remains nearly unperturbed even in the presence of losses and disorder in the component values in contrast to the quasi-BIC corner state Fig.~\ref{fig:Circuit}e which strongly hybridizes with the bulk states Fig.~\ref{fig:Circuit}d.

The above robustness is especially interesting since the fluctuations in the values of capacitors in the circuit simultaneously induce off-diagonal and diagonal disorder. Nevertheless, experimental results demonstrate excellent agreement with the theoretical predictions even in the presence of disorder and dissipation for system size as small as $9 \times 9$ sites highlighting the potential of higher-order topological states for constructing small-scale photonic and electronic devices. Moreover, we prove the topological origin of the observed corner state, retrieving the topological invariant from the experimental results as described in Supplementary Note~7.

To conclude, we have demonstrated the crucial role played by the next-nearest-neighbor interaction in the formation of higher-order topological states in $D_4$-symmetric systems. While the conventional 2D SSH model is gapless at zero energy, even small interaction of the next nearest neighbors opens the topological gap. Thus, our results provide a clear physical interpretation of the corner states observed in recent experiments with the arrays of microwave resonators~\cite{2019_Xie, 2019_Chen}. Furthermore, our study reveals the fundamental role of long-range interactions in the formation of higher-order topological phases and highlights the potential of resonant electric circuits to design and test novel topological structures.


\section*{\small Methods}

{\scriptsize

\textbf{Tight-binding model}

To find the dispersion of bulk modes, we construct the Bloch Hamiltonian which is defined in the reciprocal space for a unit cell including four sites and describes bulk excitations in the considered system. For the unit cell choice with intra-cell couplings $J$ shown in Fig.~\ref{fig:Model}c, the Bloch Hamiltonian matrix takes the following form:
\begin{equation}
    \hat{H}(k)= -
        \left(
    \begin{tabular}{c c c c}
    0 & $J+K{\rm e}^{-ik_{\rm x}}$ & $M{\rm e}^{ik_{\rm y} - ik_{\rm x}}$  & $J+K {\rm e}^{i k_{\rm y}}$ \\
        $J+K {\rm e}^{i k_{\rm x}}$ & 0 & $J+K{\rm e}^{i k_{\rm y}}$ & $M{\rm e}^{ik_{\rm y} + ik_{\rm x}}$ \\
       $M{\rm e}^{-ik_{\rm y} + ik_{\rm x}}$ & $J+K{\rm e}^{-i k_{\rm y}}$ & 0 & $J+K{\rm e}^{ik_{\rm x}}$ \\
        $J+K{\rm e}^{-ik_{\rm y}}$ & $M{\rm e}^{-ik_{\rm y} - ik_{\rm x}}$ & $J+K{\rm e}^{-ik_{\rm x}}$ & 0
    \end{tabular}
    \right),
    \label{eq:Bloch_Hamiltonian}
\end{equation}
with wave vector components $k_{\rm x}$, $k_{\rm y}$ spanning the range $[-\pi,\pi]$ and directed along the $x$- and $y$-axes shown in Fig.~\ref{fig:Model}. In the above matrix, columns and rows correspond to sites $A$, $B$, $C$, and $D$ left to right and up to down, respectively. Then, we construct a secular equation det$|\hat{H}(k)-\varepsilon\hat{I}|=0$, $\hat{I}$ being the $4 \times 4$ unity matrix, which yields four solutions for eigenvalues $\varepsilon(k)$ describing the dispersion of four bulk bands. As shown in Supplementary Note 1, three of these bands are located above zero-energy bandgap, while one band remains below the bandgap. Retrieving the topological invariant from experimental data, we focus on this isolated band. The calculated dispersion diagram is depicted in Fig.~\ref{fig:Eigenmodes}h.

\ \

\textbf{Topological invariant calculation}

To explore the topological properties of our $D_4$-symmetric model Fig.~\ref{fig:Model}c, we apply the technique of Ref.~\cite{2019_Benalcazar} suitable for systems with $C_{n}$ rotational symmetry. To this end, we introduce the matrix of  rotation operator by the angle $\pi/2$ that swaps the sites of the unit cell:
\begin{equation}
    \hat{R}_{4}= 
    \begin{pmatrix}
        0 \ 1 \  0 \ 0\\ 0 \ 0 \ 1 \ 0 \\ 0 \ 0 \ 0 \ 1 \\ 1 \ 0 \ 0 \ 0
    \end{pmatrix},
\end{equation}
which has the eigenvalues $e^{2\pi i (p-1)/4}$ for $p=1,2,3,4$ describing different behavior of the eigenvector under symmetry transformation. $C_4$ symmetry transformation brings $\Gamma$-point with coordinates $(k_{x},k_{y}) = (0,0)$ and $M$-point with $(k_x,k_y) = (\pi,\pi)$ to the equivalent points of reciprocal space. Hence, as it is straightforward to check, $[\hat{H}(\Gamma),\hat{R}_4]=0$ and  $[\hat{H}(M),\hat{R}_4]=0$. As a result, the eigenstates of the Hamiltonians $\hat{H}(\Gamma)$ and $\hat{H}(M)$ can be enumerated by the index $p$, related to the eigenvalues of $C_4$ rotation operator.

Calculating the topological invariant, we also exploit the rotation by the angle $\pi$ described by the operator $\hat{R}_2=\hat{R}_4^2$. This transformation commutes not only with $\hat{H}(\Gamma)$ and $\hat{H}(M)$, but also with $\hat{H}(X)$, where $X$ point of the Brillouin zone has the coordinates $(k_x,k_y) = (\pi,0)$. Accordingly, we label the eigenstates of the Hamiltonian $\hat{H}(X)$ by the eigenvalues of $C_2$ rotation operator.

The topological invariant is constructed by tracking the number of eigenstates with a certain law of transformation (i.e. fixed index $p$) below the bandgap~\cite{2019_Benalcazar}:
\begin{equation}
    \chi^{(4)}\equiv 
\begin{pmatrix}
    \#X_1^{(2)}-\#\Gamma_1^{(2)}\\
    \#M_1^{(4)}-\#\Gamma_1^{(4)}\\
    \#M_2^{(4)}-\#\Gamma_2^{(4)}
    \end{pmatrix}\:.
    \label{eq:chi_Definition}
\end{equation}
Here, the upper indices $(2)$ and $(4)$ correspond to $\hat{R}_{2}$ and $\hat{R}_{4}$ operators, respectively, lower indices denote the value of $p$ for the rotation operator eigenvalues and the symbol $\#$ in front of the high-symmetry point defines the number of eigenfunctions with a given transformation law below the bandgap.

As further discussed in Supplementary Note 2, if the unit cell is chosen with weaker $J$ links inside, the topological invariant is equal to $\chi^{(4)}=(-1,-1,0)$. On the other hand, choosing the unit cell with $K$ and $M$ links inside, we obtain $\tilde{\chi}^{(4)}=(0,0,0)$. These results indicate that the topological corner state arises only at the weak link corner of our system.

\ \

\textbf{Generalized chiral symmetry}

The energies and the eigenstates of the four bulk bands are linked to each other via so-called generalized chiral symmetry described by the operator $\hat{\Gamma}_4$
\begin{equation}
   \Gamma_4 = 
    \begin{pmatrix}
    1 & 0 & 0 & 0 \\
    0 & i & 0 & 0 \\
    0 & 0 & -1 & 0 \\
    0 & 0 & 0 & -i 
    \end{pmatrix}.
\end{equation}
Applying this operator to the Bloch Hamiltonian Eq.~\eqref{eq:Bloch_Hamiltonian} several times, we obtain a set of matrices:

\begin{equation}
    \begin{gathered}
        \hat{\Gamma}_{4}\hat{H}(k)\hat{\Gamma}_{4}^{-1} = \hat{H}_{\rm 1}(k), \\ 
        \hat{\Gamma}_{4}\hat{H}_{\rm 1}(k)\hat{\Gamma}_{4}^{-1} = \hat{H}_{\rm 2}(k), \\ 
        \hat{\Gamma}_{4}\hat{H}_{\rm 2}(k)\hat{\Gamma}_{4}^{-1} = \hat{H}_{\rm 3}(k), \\ 
        \hat{H}(k) + \hat{H}_{\rm 1}(k) + \hat{H}_{\rm 2}(k) + \hat{H}_{\rm 3}(k) = 0,
    \end{gathered}
    \label{eq:Symmetry}
\end{equation}

By the construction, the traces of all introduced matrices are equal: ${\rm Tr}\{\hat{H}_{\rm 0}\}= {\rm Tr}\{\hat{\Gamma}_{4}\hat{H}_{\rm 0}\hat{\Gamma}_{4}^{-1}\} = {\rm Tr}\{\hat{H}_{\rm 1}\} = {\rm Tr}\{\hat{H}_{\rm 2}\} = {\rm Tr}\{\hat{H}_{\rm 3}\}$. On the other hand, since the sum of the matrices is zero, ${\rm Tr}\{\hat{H} +  \hat{H}_{\rm 1} + \hat{H}_{\rm 2} + \hat{H}_{\rm 3}\}=0$. Thus, the trace of the Hamiltonian $\hat{H}$ vanishes:

\begin{equation*}
    {\rm Tr}\{\hat{H}\} = 0.
\end{equation*}

As a result, the sum of the four eigenvalues of Bloch Hamiltonian for the given $(k_x,k_y)$ is equal to zero, which is seen at Fig.~\ref{fig:Eigenmodes}h, while all eigenstates can be restored from the single eigenstate $\ket{\psi}$ applying generalized chiral symmetry operator: $\ket{\psi_1}=\hat{\Gamma}_4\,\ket{\psi}$, $\ket{\psi_2}=\hat{\Gamma}_4\,\ket{\psi_1}$ and $\ket{\psi_3}=\hat{\Gamma}_4\,\ket{\psi_2}$.

\ \

\textbf{Electric circuit realization}

To construct the electric circuit implementing the proposed model, we start from the explicit form of tight-binding problem Eq.(\ref{eq:Tight_binding}), considering bulk node $(m,n)$ labelled with index $B$ in Fig.~\ref{fig:Model}c as an example:
\begin{multline}
    -J(\beta_{m-1,n}+\beta_{m,n-1}) - K(\beta_{m,n+1}+\beta_{m+1,n}) -\\
    - M\beta_{m+1,n+1} = \varepsilon\beta_{mn}.
    \label{eq:TB_Site}
\end{multline}
At the same time, potentials and current in the electric circuit are governed by Kirchhoff's rules $\sum_{m',n'}I_{m'n',mn}=0$, stating that the sum of all currents flowing into an arbitrary node $(m,n)$ from all of its neighbors $(m',n')$ vanishes. The second Kirchhoff's rule is satisfied automatically by introducing on-site time-dependent potentials $\varphi_{mn}$.

Next, we introduce frequency-dependent complex admittances of the links following the time convention $e^{-i\omega t}$ for varying fields for consistency with Schr{\"o}dinger equation describing the tight-binding model. With these conventions, the admittances read $\sigma_{C_{\rm J}}(\omega) = -i\omega C_{\rm J}$, $\sigma_{C_{\rm K}}(\omega) = -i\omega C_{\rm K}$, $\sigma_{C_{\rm M}}(\omega) = -i\omega C_{\rm M}$ and $\sigma_{L}(\omega) = -1/(i\omega L)$. 

For the corresponding node $(B)$ in the circuit [Fig.~\ref{fig:Circuit}a] Kirchhoff's current rule combined with Ohm's law $I_{m'n',mn} = \sigma(\omega)(\varphi_{m'n'}-\varphi_{mn})$ reads $-\sigma_{C_{\rm J}}(\varphi_{m-1,n}+\varphi_{m,n-1}) - \sigma_{C_{\rm K}}(\varphi_{m,n+1}+\varphi_{m+1,n}) - \sigma_{C_{\rm M}}\varphi_{m+1,n+1} = (-2\sigma_{C_{\rm J}} - 2\sigma_{C_{\rm K}} - \sigma_{C_{\rm M}} - \sigma_{C_{\rm J}})\varphi_{mn}$. Dividing the above equation by $\sigma_{C_{\rm J}}$, we obtain
\begin{multline}
    -(\varphi_{m-1,n}+\varphi_{m,n-1}) - \frac{C_{\rm K}}{C_{\rm J}}(\varphi_{m,n+1}+\varphi_{m+1,n}) -\\
    - \frac{C_{\rm M}}{C_{\rm J}}\varphi_{m+1,n+1} = \left[\frac{\omega_{0}^2}{\omega^2}-\left(2+2\frac{C_{K}}{C_{J}}+\frac{C_{M}}{C_{J}}\right)\right]\,\varphi_{mn}\:,
    \label{eq:TB_Circuit}
\end{multline}
where $\omega_0^2=1/(L\,C_J)$. This equation describes on-site potential distributions $\varphi_{mn}$ for the circuit eigenmode with frequency $\omega$ and clearly resembles tight-binding problem Eq.~\eqref{eq:TB_Site}. To compensate the absence of neighbors for the boundary nodes maintaining the correspondence between Eq.~\eqref{eq:TB_Site} and Eq.~\eqref{eq:TB_Circuit}, nodes at the sides of the circuit are grounded with additional elements $C_{\rm J}$, $C_{\rm K}$, and $C_{\rm M}$ in parallel to the inductors $L$, in accordance with Fig.~\ref{fig:Circuit}a. Further details on electric circuit model, including the discussion of the boundary conditions, are provided in Supplementary Note 3.

\ \

\textbf{Experimental setup and measurements}

We implement the proposed circuit in the form of a single layer two-sided printed circuit board (PCB) made on the FR4 substrate. The circuit includes $81$ nodes arranged in $9\times 9$ lattice, as shown in Fig.~\ref{fig:Circuit}a,b. The dimensions of the PCB are $31 \times 31$ cm, and the thickness is $2$ mm. Each node of the circuit contains two MCX-type coaxial cable connectors to attach the measurement equipment. The values of circuit elements are $L=(23.19 \pm 0.04)\, \mu{\rm H}$, $C_{\rm J} = (931 \pm 3)\,{\rm nF}$, and $C_{\rm K} = C_{\rm M} = (4020 \pm 10)\,{\rm nF}$. To sort the elements up to the tolerances of $\pm 0.17\%$ for inductors $L$, $\pm 0.32\%$ for capacitors $C_{\rm J}$, and $\pm 0.25\%$ for capacitors $C_{\rm K}$ and $C_{\rm M}$, we use Mastech MS5308 LCR-meter. To characterize resonances in the circuit, we measure the frequency-dependent on-site voltage response between the given circuit node and ground when the external harmonic signal source with amplitude $U_{\rm ext}=63~{\rm mV}$ (peak-to-peak voltage $126~{\rm mV}$) and series impedance of $50~{\rm Ohm}$ is successively attached between every node of the circuit and ground. We study circuit response in the frequency range $f=(1 \dots 20)$~kHz, obtaining curves with $1000$ uniformly spaced frequency points. All $81$ voltage curves are shown in Fig.~\ref{fig:Circuit}c. Such extensive measurements allow us to plot full voltage distributions at the nodes of the circuit in the mentioned frequency range, some of which are shown in Fig.~\ref{fig:Circuit}d-g. We perform experimental studies with the help of open-source hardware platform OSA103 Mini which includes both the generator and the measurement equipment and allows automating the measurement process. To verify the results, we check the obtained voltage spectra with the help of Keithley $3390$ signal generator and Rohde\&Schwarz HMO $2022$ oscilloscope. Further details on the components used and their preparation, as well as on the equipment and measurements, are given in Supplementary Note 5.

\ \

\textbf{Numerical simulations}

We perform full numerical simulations of the extended SSH circuit with the help of Keysight Advanced Design System (ADS). Considering the same protocols as in the experimental study, we apply them to a set of circuits with broadly varied parameters of inductors and capacitors, which allows us studying the robustness of circuit resonances towards diagonal and off-diagonal disorder. Besides, we compare the effects of ohmic losses and fluctuations in element values on circuit spectrum and visualize the associated changes in profiles of characteristic resonances. Further details along with simulation results can be found in Supplementary Note 6.

\ \

\textbf{Topological invariant retrieval from experimental data}

To fully support our theoretical findings, we extract the topological invariant directly from the experimental measurements of voltage distributions in the circuit. To realize such a procedure, we drive the node $(5,5)$ with a harmonic signal at the amplitude $U_{ext} = 50~{\rm mV}$ (peak-to-peak voltage $U_{ext} = 100~{\rm mV}$) in the frequency range $f=1...20~{\rm kHz}$ and measure the induced voltage between the given node $(m,n)$ and ground at all nodes of the circuit keeping the external source located at node $(5,5)$ in contrast to spectrum measurements and maps in Fig.~\ref{fig:Circuit}. Moreover, along with voltage amplitude, we measure relative phases of voltages at all nodes taking the phase of voltage at the node $(5,5)$ as a reference. Performing the procedure outlined in Supplementary Note 7, we extract the approximated Bloch wave function of the bulk state located below the bandgap, analyzing voltage distribution in the circuit at $f=18.616~{\rm kHz}$. For the retrieved Bloch wave function, we apply the same procedure as in the theoretical calculation of topological invariant (Supplementary Note 2) and obtain consistent results proving the topological origin of the observed corner state. 

}

\subsection*{\small Acknowledgments}
Theoretical models were supported by the Russian Foundation for Basic Research (grant No.~18-29-20037), experimental studies were supported by the Russian Science Foundation (grant No.~20-72-10065). N.O. and M.G. acknowledge partial support by the Foundation for the Advancement of Theoretical Physics and Mathematics ``Basis''.

\subsection*{\small Author contributions}
M.G. and D.Z. conceived the idea. M.G. and N.O supervised the project. V.K., M.G., N.O. and D.Z. developed the theoretical models. N.O. performed numerical studies. A.R. and N.O. carried out circuit simulations. N.O., M.G. and P.S. developed the circuit model. A.D., A.R., O.B. and N.O. fabricated the experimental setup. O.B., A.D. and P.S. performed the experiments. A.R. and N.O. processed the experimental results. N.O. and M.G. prepared the paper with the input from all other authors.

\subsection*{\small Data availability}
The data that support the findings of this study are available from the corresponding authors upon request.

\subsection*{\small Competing interests}
The authors declare that they have no competing interests.

\subsection*{\small Additional information}
Correspondence and requests for materials should be addressed to M.G. (email: m.gorlach@metalab.ifmo.ru) or N.O. (email: nikita.olekhno@metalab.ifmo.ru).


\bibliography{D4_BibFile}

\begin{thebibliography}{32}%
\makeatletter
\providecommand \@ifxundefined [1]{%
 \@ifx{#1\undefined}
}%
\providecommand \@ifnum [1]{%
 \ifnum #1\expandafter \@firstoftwo
 \else \expandafter \@secondoftwo
 \fi
}%
\providecommand \@ifx [1]{%
 \ifx #1\expandafter \@firstoftwo
 \else \expandafter \@secondoftwo
 \fi
}%
\providecommand \natexlab [1]{#1}%
\providecommand \enquote  [1]{``#1''}%
\providecommand \bibnamefont  [1]{#1}%
\providecommand \bibfnamefont [1]{#1}%
\providecommand \citenamefont [1]{#1}%
\providecommand \href@noop [0]{\@secondoftwo}%
\providecommand \href [0]{\begingroup \@sanitize@url \@href}%
\providecommand \@href[1]{\@@startlink{#1}\@@href}%
\providecommand \@@href[1]{\endgroup#1\@@endlink}%
\providecommand \@sanitize@url [0]{\catcode `\\12\catcode `\$12\catcode
  `\&12\catcode `\#12\catcode `\^12\catcode `\_12\catcode `\%12\relax}%
\providecommand \@@startlink[1]{}%
\providecommand \@@endlink[0]{}%
\providecommand \url  [0]{\begingroup\@sanitize@url \@url }%
\providecommand \@url [1]{\endgroup\@href {#1}{\urlprefix }}%
\providecommand \urlprefix  [0]{URL }%
\providecommand \Eprint [0]{\href }%
\providecommand \doibase [0]{http://dx.doi.org/}%
\providecommand \selectlanguage [0]{\@gobble}%
\providecommand \bibinfo  [0]{\@secondoftwo}%
\providecommand \bibfield  [0]{\@secondoftwo}%
\providecommand \translation [1]{[#1]}%
\providecommand \BibitemOpen [0]{}%
\providecommand \bibitemStop [0]{}%
\providecommand \bibitemNoStop [0]{.\EOS\space}%
\providecommand \EOS [0]{\spacefactor3000\relax}%
\providecommand \BibitemShut  [1]{\csname bibitem#1\endcsname}%
\let\auto@bib@innerbib\@empty
\bibitem [{\citenamefont {Benalcazar}\ \emph {et~al.}(2017)\citenamefont
  {Benalcazar}, \citenamefont {Bernevig},\ and\ \citenamefont
  {Hughes}}]{2017_Benalcazar_Science}%
  \BibitemOpen
  \bibfield  {author} {\bibinfo {author} {\bibfnamefont {Wladimir~A.}\
  \bibnamefont {Benalcazar}}, \bibinfo {author} {\bibfnamefont {B.~Andrei}\
  \bibnamefont {Bernevig}}, \ and\ \bibinfo {author} {\bibfnamefont
  {Taylor~L.}\ \bibnamefont {Hughes}},\ }\bibfield  {title} {\enquote {\bibinfo
  {title} {Quantized electric multipole insulators},}\ }\href {\doibase
  10.1126/science.aah6442} {\bibfield  {journal} {\bibinfo  {journal}
  {Science}\ }\textbf {\bibinfo {volume} {357}},\ \bibinfo {pages} {61--66}
  (\bibinfo {year} {2017})}\BibitemShut {NoStop}%
\bibitem [{\citenamefont {Schindler}\ \emph
  {et~al.}(2018{\natexlab{a}})\citenamefont {Schindler}, \citenamefont {Cook},
  \citenamefont {Vergniory}, \citenamefont {Wang}, \citenamefont {Parkin},
  \citenamefont {Bernevig},\ and\ \citenamefont {Neupert}}]{2018_Schindler}%
  \BibitemOpen
  \bibfield  {author} {\bibinfo {author} {\bibfnamefont {Frank}\ \bibnamefont
  {Schindler}}, \bibinfo {author} {\bibfnamefont {Ashley~M.}\ \bibnamefont
  {Cook}}, \bibinfo {author} {\bibfnamefont {Maia~G.}\ \bibnamefont
  {Vergniory}}, \bibinfo {author} {\bibfnamefont {Zhijun}\ \bibnamefont
  {Wang}}, \bibinfo {author} {\bibfnamefont {Stuart S.~P.}\ \bibnamefont
  {Parkin}}, \bibinfo {author} {\bibfnamefont {B.~Andrei}\ \bibnamefont
  {Bernevig}}, \ and\ \bibinfo {author} {\bibfnamefont {Titus}\ \bibnamefont
  {Neupert}},\ }\bibfield  {title} {\enquote {\bibinfo {title} {Higher-order
  topological insulators},}\ }\href {\doibase 10.1126/sciadv.aat0346}
  {\bibfield  {journal} {\bibinfo  {journal} {Science Advances}\ }\textbf
  {\bibinfo {volume} {4}},\ \bibinfo {pages} {eaat0346} (\bibinfo {year}
  {2018}{\natexlab{a}})}\BibitemShut {NoStop}%
\bibitem [{\citenamefont {Xue}\ \emph {et~al.}(2018)\citenamefont {Xue},
  \citenamefont {Yang}, \citenamefont {Gao}, \citenamefont {Chong},\ and\
  \citenamefont {Zhang}}]{2018_Xue}%
  \BibitemOpen
  \bibfield  {author} {\bibinfo {author} {\bibfnamefont {Haoran}\ \bibnamefont
  {Xue}}, \bibinfo {author} {\bibfnamefont {Yahui}\ \bibnamefont {Yang}},
  \bibinfo {author} {\bibfnamefont {Fei}\ \bibnamefont {Gao}}, \bibinfo
  {author} {\bibfnamefont {Yidong}\ \bibnamefont {Chong}}, \ and\ \bibinfo
  {author} {\bibfnamefont {Baile}\ \bibnamefont {Zhang}},\ }\bibfield  {title}
  {\enquote {\bibinfo {title} {Acoustic higher-order topological insulator on a
  kagome lattice},}\ }\href {\doibase 10.1038/s41563-018-0251-x} {\bibfield
  {journal} {\bibinfo  {journal} {Nature Materials}\ }\textbf {\bibinfo
  {volume} {18}},\ \bibinfo {pages} {108--112} (\bibinfo {year}
  {2018})}\BibitemShut {NoStop}%
\bibitem [{\citenamefont {Ni}\ \emph {et~al.}(2018)\citenamefont {Ni},
  \citenamefont {Weiner}, \citenamefont {Al{\`{u}}},\ and\ \citenamefont
  {Khanikaev}}]{2018_Ni_NatureMat}%
  \BibitemOpen
  \bibfield  {author} {\bibinfo {author} {\bibfnamefont {Xiang}\ \bibnamefont
  {Ni}}, \bibinfo {author} {\bibfnamefont {Matthew}\ \bibnamefont {Weiner}},
  \bibinfo {author} {\bibfnamefont {Andrea}\ \bibnamefont {Al{\`{u}}}}, \ and\
  \bibinfo {author} {\bibfnamefont {Alexander~B.}\ \bibnamefont {Khanikaev}},\
  }\bibfield  {title} {\enquote {\bibinfo {title} {Observation of higher-order
  topological acoustic states protected by generalized chiral symmetry},}\
  }\href {\doibase 10.1038/s41563-018-0252-9} {\bibfield  {journal} {\bibinfo
  {journal} {Nature Materials}\ }\textbf {\bibinfo {volume} {18}},\ \bibinfo
  {pages} {113--120} (\bibinfo {year} {2018})}\BibitemShut {NoStop}%
\bibitem [{\citenamefont {Wu}\ and\ \citenamefont {Hu}(2015)}]{2015_Wu}%
  \BibitemOpen
  \bibfield  {author} {\bibinfo {author} {\bibfnamefont {Long-Hua}\
  \bibnamefont {Wu}}\ and\ \bibinfo {author} {\bibfnamefont {Xiao}\
  \bibnamefont {Hu}},\ }\bibfield  {title} {\enquote {\bibinfo {title} {{Scheme
  for Achieving a Topological Photonic Crystal by Using Dielectric
  Material}},}\ }\href {\doibase 10.1103/physrevlett.114.223901} {\bibfield
  {journal} {\bibinfo  {journal} {Physical Review Letters}\ }\textbf {\bibinfo
  {volume} {114}},\ \bibinfo {pages} {223901} (\bibinfo {year}
  {2015})}\BibitemShut {NoStop}%
\bibitem [{\citenamefont {Schindler}\ \emph
  {et~al.}(2018{\natexlab{b}})\citenamefont {Schindler}, \citenamefont {Wang},
  \citenamefont {Vergniory}, \citenamefont {Cook}, \citenamefont {Murani},
  \citenamefont {Sengupta}, \citenamefont {Kasumov}, \citenamefont {Deblock},
  \citenamefont {Jeon}, \citenamefont {Drozdov}, \citenamefont {Bouchiat},
  \citenamefont {Gu{\'{e}}ron}, \citenamefont {Yazdani}, \citenamefont
  {Bernevig},\ and\ \citenamefont {Neupert}}]{2018_Schindler_Bismuth}%
  \BibitemOpen
  \bibfield  {author} {\bibinfo {author} {\bibfnamefont {Frank}\ \bibnamefont
  {Schindler}}, \bibinfo {author} {\bibfnamefont {Zhijun}\ \bibnamefont
  {Wang}}, \bibinfo {author} {\bibfnamefont {Maia~G.}\ \bibnamefont
  {Vergniory}}, \bibinfo {author} {\bibfnamefont {Ashley~M.}\ \bibnamefont
  {Cook}}, \bibinfo {author} {\bibfnamefont {Anil}\ \bibnamefont {Murani}},
  \bibinfo {author} {\bibfnamefont {Shamashis}\ \bibnamefont {Sengupta}},
  \bibinfo {author} {\bibfnamefont {Alik~Yu.}\ \bibnamefont {Kasumov}},
  \bibinfo {author} {\bibfnamefont {Richard}\ \bibnamefont {Deblock}}, \bibinfo
  {author} {\bibfnamefont {Sangjun}\ \bibnamefont {Jeon}}, \bibinfo {author}
  {\bibfnamefont {Ilya}\ \bibnamefont {Drozdov}}, \bibinfo {author}
  {\bibfnamefont {H{\'{e}}l{\`{e}}ne}\ \bibnamefont {Bouchiat}}, \bibinfo
  {author} {\bibfnamefont {Sophie}\ \bibnamefont {Gu{\'{e}}ron}}, \bibinfo
  {author} {\bibfnamefont {Ali}\ \bibnamefont {Yazdani}}, \bibinfo {author}
  {\bibfnamefont {B.~Andrei}\ \bibnamefont {Bernevig}}, \ and\ \bibinfo
  {author} {\bibfnamefont {Titus}\ \bibnamefont {Neupert}},\ }\bibfield
  {title} {\enquote {\bibinfo {title} {Higher-order topology in bismuth},}\
  }\href {\doibase 10.1038/s41567-018-0224-7} {\bibfield  {journal} {\bibinfo
  {journal} {Nature Physics}\ }\textbf {\bibinfo {volume} {14}},\ \bibinfo
  {pages} {918--924} (\bibinfo {year} {2018}{\natexlab{b}})}\BibitemShut
  {NoStop}%
\bibitem [{\citenamefont {Serra-Garcia}\ \emph {et~al.}(2018)\citenamefont
  {Serra-Garcia}, \citenamefont {Peri}, \citenamefont {S\"{u}sstrunk},
  \citenamefont {Bilal}, \citenamefont {Larsen}, \citenamefont {Villanueva},\
  and\ \citenamefont {Huber}}]{2018_Serra_Garcia}%
  \BibitemOpen
  \bibfield  {author} {\bibinfo {author} {\bibfnamefont {Marc}\ \bibnamefont
  {Serra-Garcia}}, \bibinfo {author} {\bibfnamefont {Valerio}\ \bibnamefont
  {Peri}}, \bibinfo {author} {\bibfnamefont {Roman}\ \bibnamefont
  {S\"{u}sstrunk}}, \bibinfo {author} {\bibfnamefont {Osama~R.}\ \bibnamefont
  {Bilal}}, \bibinfo {author} {\bibfnamefont {Tom}\ \bibnamefont {Larsen}},
  \bibinfo {author} {\bibfnamefont {Luis~Guillermo}\ \bibnamefont
  {Villanueva}}, \ and\ \bibinfo {author} {\bibfnamefont {Sebastian~D.}\
  \bibnamefont {Huber}},\ }\bibfield  {title} {\enquote {\bibinfo {title}
  {Observation of a phononic quadrupole topological insulator},}\ }\href
  {\doibase 10.1038/nature25156} {\bibfield  {journal} {\bibinfo  {journal}
  {Nature}\ }\textbf {\bibinfo {volume} {555}},\ \bibinfo {pages} {342--345}
  (\bibinfo {year} {2018})}\BibitemShut {NoStop}%
\bibitem [{\citenamefont {Mittal}\ \emph {et~al.}(2019)\citenamefont {Mittal},
  \citenamefont {Orre}, \citenamefont {Zhu}, \citenamefont {Gorlach},
  \citenamefont {Poddubny},\ and\ \citenamefont {Hafezi}}]{2019_Mittal}%
  \BibitemOpen
  \bibfield  {author} {\bibinfo {author} {\bibfnamefont {Sunil}\ \bibnamefont
  {Mittal}}, \bibinfo {author} {\bibfnamefont {Venkata~Vikram}\ \bibnamefont
  {Orre}}, \bibinfo {author} {\bibfnamefont {Guanyu}\ \bibnamefont {Zhu}},
  \bibinfo {author} {\bibfnamefont {Maxim~A.}\ \bibnamefont {Gorlach}},
  \bibinfo {author} {\bibfnamefont {Alexander}\ \bibnamefont {Poddubny}}, \
  and\ \bibinfo {author} {\bibfnamefont {Mohammad}\ \bibnamefont {Hafezi}},\
  }\bibfield  {title} {\enquote {\bibinfo {title} {Photonic quadrupole
  topological phases},}\ }\href {\doibase 10.1038/s41566-019-0452-0} {\bibfield
   {journal} {\bibinfo  {journal} {Nature Photonics}\ }\textbf {\bibinfo
  {volume} {13}},\ \bibinfo {pages} {692--696} (\bibinfo {year}
  {2019})}\BibitemShut {NoStop}%
\bibitem [{\citenamefont {Hassan}\ \emph {et~al.}(2019)\citenamefont {Hassan},
  \citenamefont {Kunst}, \citenamefont {Moritz}, \citenamefont {Andler},
  \citenamefont {Bergholtz},\ and\ \citenamefont {Bourennane}}]{2019_Hassan}%
  \BibitemOpen
  \bibfield  {author} {\bibinfo {author} {\bibfnamefont {Ashraf~El}\
  \bibnamefont {Hassan}}, \bibinfo {author} {\bibfnamefont {Flore~K.}\
  \bibnamefont {Kunst}}, \bibinfo {author} {\bibfnamefont {Alexander}\
  \bibnamefont {Moritz}}, \bibinfo {author} {\bibfnamefont {Guillermo}\
  \bibnamefont {Andler}}, \bibinfo {author} {\bibfnamefont {Emil~J.}\
  \bibnamefont {Bergholtz}}, \ and\ \bibinfo {author} {\bibfnamefont {Mohamed}\
  \bibnamefont {Bourennane}},\ }\bibfield  {title} {\enquote {\bibinfo {title}
  {Corner states of light in photonic waveguides},}\ }\href {\doibase
  10.1038/s41566-019-0519-y} {\bibfield  {journal} {\bibinfo  {journal} {Nature
  Photonics}\ }\textbf {\bibinfo {volume} {13}},\ \bibinfo {pages} {697--700}
  (\bibinfo {year} {2019})}\BibitemShut {NoStop}%
\bibitem [{\citenamefont {Peterson}\ \emph {et~al.}(2018)\citenamefont
  {Peterson}, \citenamefont {Benalcazar}, \citenamefont {Hughes},\ and\
  \citenamefont {Bahl}}]{2018_Peterson}%
  \BibitemOpen
  \bibfield  {author} {\bibinfo {author} {\bibfnamefont {Christopher~W.}\
  \bibnamefont {Peterson}}, \bibinfo {author} {\bibfnamefont {Wladimir~A.}\
  \bibnamefont {Benalcazar}}, \bibinfo {author} {\bibfnamefont {Taylor~L.}\
  \bibnamefont {Hughes}}, \ and\ \bibinfo {author} {\bibfnamefont {Gaurav}\
  \bibnamefont {Bahl}},\ }\bibfield  {title} {\enquote {\bibinfo {title} {A
  quantized microwave quadrupole insulator with topologically protected corner
  states},}\ }\href {\doibase 10.1038/nature25777} {\bibfield  {journal}
  {\bibinfo  {journal} {Nature}\ }\textbf {\bibinfo {volume} {555}},\ \bibinfo
  {pages} {346--350} (\bibinfo {year} {2018})}\BibitemShut {NoStop}%
\bibitem [{\citenamefont {Li}\ \emph {et~al.}(2020)\citenamefont {Li},
  \citenamefont {Zhirihin}, \citenamefont {Gorlach}, \citenamefont {Ni},
  \citenamefont {Filonov}, \citenamefont {Slobozhanyuk}, \citenamefont
  {Al{\`{u}}},\ and\ \citenamefont {Khanikaev}}]{2020_Li}%
  \BibitemOpen
  \bibfield  {author} {\bibinfo {author} {\bibfnamefont {Mengyao}\ \bibnamefont
  {Li}}, \bibinfo {author} {\bibfnamefont {Dmitry}\ \bibnamefont {Zhirihin}},
  \bibinfo {author} {\bibfnamefont {Maxim}\ \bibnamefont {Gorlach}}, \bibinfo
  {author} {\bibfnamefont {Xiang}\ \bibnamefont {Ni}}, \bibinfo {author}
  {\bibfnamefont {Dmitry}\ \bibnamefont {Filonov}}, \bibinfo {author}
  {\bibfnamefont {Alexey}\ \bibnamefont {Slobozhanyuk}}, \bibinfo {author}
  {\bibfnamefont {Andrea}\ \bibnamefont {Al{\`{u}}}}, \ and\ \bibinfo {author}
  {\bibfnamefont {Alexander~B.}\ \bibnamefont {Khanikaev}},\ }\bibfield
  {title} {\enquote {\bibinfo {title} {Higher-order topological states in
  photonic kagome crystals with long-range interactions},}\ }\href {\doibase
  10.1038/s41566-019-0561-9} {\bibfield  {journal} {\bibinfo  {journal} {Nature
  Photonics}\ }\textbf {\bibinfo {volume} {14}},\ \bibinfo {pages} {89--94}
  (\bibinfo {year} {2020})}\BibitemShut {NoStop}%
\bibitem [{\citenamefont {Imhof}\ \emph {et~al.}(2018)\citenamefont {Imhof},
  \citenamefont {Berger}, \citenamefont {Bayer}, \citenamefont {Brehm},
  \citenamefont {Molenkamp}, \citenamefont {Kiessling}, \citenamefont
  {Schindler}, \citenamefont {Lee}, \citenamefont {Greiter}, \citenamefont
  {Neupert},\ and\ \citenamefont {Thomale}}]{2018_Imhof}%
  \BibitemOpen
  \bibfield  {author} {\bibinfo {author} {\bibfnamefont {Stefan}\ \bibnamefont
  {Imhof}}, \bibinfo {author} {\bibfnamefont {Christian}\ \bibnamefont
  {Berger}}, \bibinfo {author} {\bibfnamefont {Florian}\ \bibnamefont {Bayer}},
  \bibinfo {author} {\bibfnamefont {Johannes}\ \bibnamefont {Brehm}}, \bibinfo
  {author} {\bibfnamefont {Laurens~W.}\ \bibnamefont {Molenkamp}}, \bibinfo
  {author} {\bibfnamefont {Tobias}\ \bibnamefont {Kiessling}}, \bibinfo
  {author} {\bibfnamefont {Frank}\ \bibnamefont {Schindler}}, \bibinfo {author}
  {\bibfnamefont {Ching~Hua}\ \bibnamefont {Lee}}, \bibinfo {author}
  {\bibfnamefont {Martin}\ \bibnamefont {Greiter}}, \bibinfo {author}
  {\bibfnamefont {Titus}\ \bibnamefont {Neupert}}, \ and\ \bibinfo {author}
  {\bibfnamefont {Ronny}\ \bibnamefont {Thomale}},\ }\bibfield  {title}
  {\enquote {\bibinfo {title} {Topolectrical-circuit realization of topological
  corner modes},}\ }\href {\doibase 10.1038/s41567-018-0246-1} {\bibfield
  {journal} {\bibinfo  {journal} {Nature Physics}\ }\textbf {\bibinfo {volume}
  {14}},\ \bibinfo {pages} {925--929} (\bibinfo {year} {2018})}\BibitemShut
  {NoStop}%
\bibitem [{\citenamefont {Serra-Garcia}\ \emph {et~al.}(2019)\citenamefont
  {Serra-Garcia}, \citenamefont {S\"{u}sstrunk},\ and\ \citenamefont
  {Huber}}]{2019_Serra_Garcia}%
  \BibitemOpen
  \bibfield  {author} {\bibinfo {author} {\bibfnamefont {Marc}\ \bibnamefont
  {Serra-Garcia}}, \bibinfo {author} {\bibfnamefont {Roman}\ \bibnamefont
  {S\"{u}sstrunk}}, \ and\ \bibinfo {author} {\bibfnamefont {Sebastian~D.}\
  \bibnamefont {Huber}},\ }\bibfield  {title} {\enquote {\bibinfo {title}
  {Observation of quadrupole transitions and edge mode topology in an {LC}
  circuit network},}\ }\href {\doibase 10.1103/physrevb.99.020304} {\bibfield
  {journal} {\bibinfo  {journal} {Physical Review B}\ }\textbf {\bibinfo
  {volume} {99}},\ \bibinfo {pages} {020304(R)} (\bibinfo {year}
  {2019})}\BibitemShut {NoStop}%
\bibitem [{\citenamefont {Bahari}\ \emph {et~al.}(2017)\citenamefont {Bahari},
  \citenamefont {Ndao}, \citenamefont {Vallini}, \citenamefont {Amili},
  \citenamefont {Fainman},\ and\ \citenamefont {Kant{\'{e}}}}]{2017_Bahari}%
  \BibitemOpen
  \bibfield  {author} {\bibinfo {author} {\bibfnamefont {Babak}\ \bibnamefont
  {Bahari}}, \bibinfo {author} {\bibfnamefont {Abdoulaye}\ \bibnamefont
  {Ndao}}, \bibinfo {author} {\bibfnamefont {Felipe}\ \bibnamefont {Vallini}},
  \bibinfo {author} {\bibfnamefont {Abdelkrim~El}\ \bibnamefont {Amili}},
  \bibinfo {author} {\bibfnamefont {Yeshaiahu}\ \bibnamefont {Fainman}}, \ and\
  \bibinfo {author} {\bibfnamefont {Boubacar}\ \bibnamefont {Kant{\'{e}}}},\
  }\bibfield  {title} {\enquote {\bibinfo {title} {Nonreciprocal lasing in
  topological cavities of arbitrary geometries},}\ }\href {\doibase
  10.1126/science.aao4551} {\bibfield  {journal} {\bibinfo  {journal}
  {Science}\ }\textbf {\bibinfo {volume} {358}},\ \bibinfo {pages} {636--640}
  (\bibinfo {year} {2017})}\BibitemShut {NoStop}%
\bibitem [{\citenamefont {Zhang}\ \emph {et~al.}(2020)\citenamefont {Zhang},
  \citenamefont {Xie}, \citenamefont {Hao}, \citenamefont {Dang}, \citenamefont
  {Xiao}, \citenamefont {Shi}, \citenamefont {Ni}, \citenamefont {Niu},
  \citenamefont {Wang}, \citenamefont {Jin}, \citenamefont {Zhang},\ and\
  \citenamefont {Xu}}]{2020_Zhang}%
  \BibitemOpen
  \bibfield  {author} {\bibinfo {author} {\bibfnamefont {Weixuan}\ \bibnamefont
  {Zhang}}, \bibinfo {author} {\bibfnamefont {Xin}\ \bibnamefont {Xie}},
  \bibinfo {author} {\bibfnamefont {Huiming}\ \bibnamefont {Hao}}, \bibinfo
  {author} {\bibfnamefont {Jianchen}\ \bibnamefont {Dang}}, \bibinfo {author}
  {\bibfnamefont {Shan}\ \bibnamefont {Xiao}}, \bibinfo {author} {\bibfnamefont
  {Shushu}\ \bibnamefont {Shi}}, \bibinfo {author} {\bibfnamefont {Haiqiao}\
  \bibnamefont {Ni}}, \bibinfo {author} {\bibfnamefont {Zhichuan}\ \bibnamefont
  {Niu}}, \bibinfo {author} {\bibfnamefont {Can}\ \bibnamefont {Wang}},
  \bibinfo {author} {\bibfnamefont {Kuijuan}\ \bibnamefont {Jin}}, \bibinfo
  {author} {\bibfnamefont {Xiangdong}\ \bibnamefont {Zhang}}, \ and\ \bibinfo
  {author} {\bibfnamefont {Xiulai}\ \bibnamefont {Xu}},\ }\bibfield  {title}
  {\enquote {\bibinfo {title} {Low-threshold topological nanolasers based on
  the second-order corner state},}\ }\href {\doibase
  10.1038/s41377-020-00352-1} {\bibfield  {journal} {\bibinfo  {journal}
  {Light: Science {\&} Applications}\ }\textbf {\bibinfo {volume} {9}},\
  \bibinfo {pages} {109} (\bibinfo {year} {2020})}\BibitemShut {NoStop}%
\bibitem [{\citenamefont {Han}\ \emph {et~al.}(2020)\citenamefont {Han},
  \citenamefont {Kang},\ and\ \citenamefont {Jeon}}]{2020_Han}%
  \BibitemOpen
  \bibfield  {author} {\bibinfo {author} {\bibfnamefont {Changhyun}\
  \bibnamefont {Han}}, \bibinfo {author} {\bibfnamefont {Minsu}\ \bibnamefont
  {Kang}}, \ and\ \bibinfo {author} {\bibfnamefont {Heonsu}\ \bibnamefont
  {Jeon}},\ }\bibfield  {title} {\enquote {\bibinfo {title} {Lasing at
  multidimensional topological states in a two-dimensional photonic crystal
  structure},}\ }\href {\doibase 10.1021/acsphotonics.0c00357} {\bibfield
  {journal} {\bibinfo  {journal} {{ACS} Photonics}\ }\textbf {\bibinfo {volume}
  {7}},\ \bibinfo {pages} {2027--2036} (\bibinfo {year} {2020})}\BibitemShut
  {NoStop}%
\bibitem [{\citenamefont {Kim}\ \emph {et~al.}(2020)\citenamefont {Kim},
  \citenamefont {Hwang}, \citenamefont {Smirnova}, \citenamefont {Jeong},
  \citenamefont {Kivshar},\ and\ \citenamefont {Park}}]{2020_Kim}%
  \BibitemOpen
  \bibfield  {author} {\bibinfo {author} {\bibfnamefont {Ha-Reem}\ \bibnamefont
  {Kim}}, \bibinfo {author} {\bibfnamefont {Min-Soo}\ \bibnamefont {Hwang}},
  \bibinfo {author} {\bibfnamefont {Daria}\ \bibnamefont {Smirnova}}, \bibinfo
  {author} {\bibfnamefont {Kwang-Yong}\ \bibnamefont {Jeong}}, \bibinfo
  {author} {\bibfnamefont {Yuri}\ \bibnamefont {Kivshar}}, \ and\ \bibinfo
  {author} {\bibfnamefont {Hong-Gyu}\ \bibnamefont {Park}},\ }\bibfield
  {title} {\enquote {\bibinfo {title} {Multipolar lasing modes from topological
  corner states},}\ }\href {\doibase 10.1038/s41467-020-19609-9} {\bibfield
  {journal} {\bibinfo  {journal} {Nature Communications}\ }\textbf {\bibinfo
  {volume} {11}},\ \bibinfo {pages} {5758} (\bibinfo {year}
  {2020})}\BibitemShut {NoStop}%
\bibitem [{\citenamefont {Chen}\ \emph {et~al.}(2019)\citenamefont {Chen},
  \citenamefont {Deng}, \citenamefont {Shi}, \citenamefont {Zhao},
  \citenamefont {Chen},\ and\ \citenamefont {Dong}}]{2019_Chen}%
  \BibitemOpen
  \bibfield  {author} {\bibinfo {author} {\bibfnamefont {Xiao-Dong}\
  \bibnamefont {Chen}}, \bibinfo {author} {\bibfnamefont {Wei-Min}\
  \bibnamefont {Deng}}, \bibinfo {author} {\bibfnamefont {Fu-Long}\
  \bibnamefont {Shi}}, \bibinfo {author} {\bibfnamefont {Fu-Li}\ \bibnamefont
  {Zhao}}, \bibinfo {author} {\bibfnamefont {Min}\ \bibnamefont {Chen}}, \ and\
  \bibinfo {author} {\bibfnamefont {Jian-Wen}\ \bibnamefont {Dong}},\
  }\bibfield  {title} {\enquote {\bibinfo {title} {Direct observation of corner
  states in second-order topological photonic crystal slabs},}\ }\href
  {\doibase 10.1103/physrevlett.122.233902} {\bibfield  {journal} {\bibinfo
  {journal} {Physical Review Letters}\ }\textbf {\bibinfo {volume} {122}},\
  \bibinfo {pages} {233902} (\bibinfo {year} {2019})}\BibitemShut {NoStop}%
\bibitem [{\citenamefont {Xie}\ \emph {et~al.}(2019)\citenamefont {Xie},
  \citenamefont {Su}, \citenamefont {Wang}, \citenamefont {Su}, \citenamefont
  {Shen}, \citenamefont {Zhan}, \citenamefont {Lu}, \citenamefont {Wang},\ and\
  \citenamefont {Chen}}]{2019_Xie}%
  \BibitemOpen
  \bibfield  {author} {\bibinfo {author} {\bibfnamefont {Bi-Ye}\ \bibnamefont
  {Xie}}, \bibinfo {author} {\bibfnamefont {Guang-Xu}\ \bibnamefont {Su}},
  \bibinfo {author} {\bibfnamefont {Hong-Fei}\ \bibnamefont {Wang}}, \bibinfo
  {author} {\bibfnamefont {Hai}\ \bibnamefont {Su}}, \bibinfo {author}
  {\bibfnamefont {Xiao-Peng}\ \bibnamefont {Shen}}, \bibinfo {author}
  {\bibfnamefont {Peng}\ \bibnamefont {Zhan}}, \bibinfo {author} {\bibfnamefont
  {Ming-Hui}\ \bibnamefont {Lu}}, \bibinfo {author} {\bibfnamefont {Zhen-Lin}\
  \bibnamefont {Wang}}, \ and\ \bibinfo {author} {\bibfnamefont {Yan-Feng}\
  \bibnamefont {Chen}},\ }\bibfield  {title} {\enquote {\bibinfo {title}
  {Visualization of higher-order topological insulating phases in
  two-dimensional dielectric photonic crystals},}\ }\href {\doibase
  10.1103/physrevlett.122.233903} {\bibfield  {journal} {\bibinfo  {journal}
  {Physical Review Letters}\ }\textbf {\bibinfo {volume} {122}},\ \bibinfo
  {pages} {233903} (\bibinfo {year} {2019})}\BibitemShut {NoStop}%
\bibitem [{\citenamefont {Liu}\ and\ \citenamefont
  {Wakabayashi}(2017)}]{2017_Wakabayashi}%
  \BibitemOpen
  \bibfield  {author} {\bibinfo {author} {\bibfnamefont {Feng}\ \bibnamefont
  {Liu}}\ and\ \bibinfo {author} {\bibfnamefont {Katsunori}\ \bibnamefont
  {Wakabayashi}},\ }\bibfield  {title} {\enquote {\bibinfo {title} {Novel
  topological phase with a zero {Berry} curvature},}\ }\href {\doibase
  10.1103/physrevlett.118.076803} {\bibfield  {journal} {\bibinfo  {journal}
  {Physical Review Letters}\ }\textbf {\bibinfo {volume} {118}},\ \bibinfo
  {pages} {076803} (\bibinfo {year} {2017})}\BibitemShut {NoStop}%
\bibitem [{\citenamefont {Thouless}(1974)}]{1974_Thouless}%
  \BibitemOpen
  \bibfield  {author} {\bibinfo {author} {\bibfnamefont {David~J.}\
  \bibnamefont {Thouless}},\ }\bibfield  {title} {\enquote {\bibinfo {title}
  {Electrons in disordered systems and the theory of localization},}\ }\href
  {\doibase https://doi.org/10.1016/0370-1573(74)90029-5} {\bibfield  {journal}
  {\bibinfo  {journal} {Physics Reports}\ }\textbf {\bibinfo {volume} {13}},\
  \bibinfo {pages} {93--142} (\bibinfo {year} {1974})}\BibitemShut {NoStop}%
\bibitem [{\citenamefont {Mukherjee}\ and\ \citenamefont
  {Rechtsman}(2020)}]{2020_Mukherjee}%
  \BibitemOpen
  \bibfield  {author} {\bibinfo {author} {\bibfnamefont {Sebabrata}\
  \bibnamefont {Mukherjee}}\ and\ \bibinfo {author} {\bibfnamefont {Mikael~C.}\
  \bibnamefont {Rechtsman}},\ }\bibfield  {title} {\enquote {\bibinfo {title}
  {Observation of {Floquet} solitons in a topological bandgap},}\ }\href
  {\doibase 10.1126/science.aba8725} {\bibfield  {journal} {\bibinfo  {journal}
  {Science}\ }\textbf {\bibinfo {volume} {368}},\ \bibinfo {pages} {856--859}
  (\bibinfo {year} {2020})}\BibitemShut {NoStop}%
\bibitem [{\citenamefont {Hsu}\ \emph {et~al.}(2016)\citenamefont {Hsu},
  \citenamefont {Zhen}, \citenamefont {Stone}, \citenamefont {Joannopoulos},\
  and\ \citenamefont {Solja{\v{c}}i{\'{c}}}}]{2016_Hsu}%
  \BibitemOpen
  \bibfield  {author} {\bibinfo {author} {\bibfnamefont {Chia~Wei}\
  \bibnamefont {Hsu}}, \bibinfo {author} {\bibfnamefont {Bo}~\bibnamefont
  {Zhen}}, \bibinfo {author} {\bibfnamefont {A.~Douglas}\ \bibnamefont
  {Stone}}, \bibinfo {author} {\bibfnamefont {John~D.}\ \bibnamefont
  {Joannopoulos}}, \ and\ \bibinfo {author} {\bibfnamefont {Marin}\
  \bibnamefont {Solja{\v{c}}i{\'{c}}}},\ }\bibfield  {title} {\enquote
  {\bibinfo {title} {Bound states in the continuum},}\ }\href {\doibase
  10.1038/natrevmats.2016.48} {\bibfield  {journal} {\bibinfo  {journal}
  {Nature Reviews Materials}\ }\textbf {\bibinfo {volume} {1}},\ \bibinfo
  {pages} {16048} (\bibinfo {year} {2016})}\BibitemShut {NoStop}%
\bibitem [{\citenamefont {Cerjan}\ \emph {et~al.}(2020)\citenamefont {Cerjan},
  \citenamefont {J\"{u}rgensen}, \citenamefont {Benalcazar}, \citenamefont
  {Mukherjee},\ and\ \citenamefont {Rechtsman}}]{2020_Cerjan}%
  \BibitemOpen
  \bibfield  {author} {\bibinfo {author} {\bibfnamefont {Alexander}\
  \bibnamefont {Cerjan}}, \bibinfo {author} {\bibfnamefont {Marius}\
  \bibnamefont {J\"{u}rgensen}}, \bibinfo {author} {\bibfnamefont
  {Wladimir~A.}\ \bibnamefont {Benalcazar}}, \bibinfo {author} {\bibfnamefont
  {Sebabrata}\ \bibnamefont {Mukherjee}}, \ and\ \bibinfo {author}
  {\bibfnamefont {Mikael~C.}\ \bibnamefont {Rechtsman}},\ }\bibfield  {title}
  {\enquote {\bibinfo {title} {Observation of a higher-order topological bound
  state in the continuum},}\ }\href {\doibase 10.1103/physrevlett.125.213901}
  {\bibfield  {journal} {\bibinfo  {journal} {Physical Review Letters}\
  }\textbf {\bibinfo {volume} {125}},\ \bibinfo {pages} {213901} (\bibinfo
  {year} {2020})}\BibitemShut {NoStop}%
\bibitem [{\citenamefont {Benalcazar}\ \emph {et~al.}(2019)\citenamefont
  {Benalcazar}, \citenamefont {Li},\ and\ \citenamefont
  {Hughes}}]{2019_Benalcazar}%
  \BibitemOpen
  \bibfield  {author} {\bibinfo {author} {\bibfnamefont {Wladimir~A.}\
  \bibnamefont {Benalcazar}}, \bibinfo {author} {\bibfnamefont {Tianhe}\
  \bibnamefont {Li}}, \ and\ \bibinfo {author} {\bibfnamefont {Taylor~L.}\
  \bibnamefont {Hughes}},\ }\bibfield  {title} {\enquote {\bibinfo {title}
  {Quantization of fractional corner charge in {$C_n$}-symmetric higher-order
  topological crystalline insulators},}\ }\href {\doibase
  10.1103/physrevb.99.245151} {\bibfield  {journal} {\bibinfo  {journal}
  {Physical Review B}\ }\textbf {\bibinfo {volume} {99}},\ \bibinfo {pages}
  {245151} (\bibinfo {year} {2019})}\BibitemShut {NoStop}%
\bibitem [{\citenamefont {Wang}\ \emph {et~al.}(2020)\citenamefont {Wang},
  \citenamefont {Price}, \citenamefont {Zhang},\ and\ \citenamefont
  {Chong}}]{2020_Wang}%
  \BibitemOpen
  \bibfield  {author} {\bibinfo {author} {\bibfnamefont {You}\ \bibnamefont
  {Wang}}, \bibinfo {author} {\bibfnamefont {Hannah~M.}\ \bibnamefont {Price}},
  \bibinfo {author} {\bibfnamefont {Baile}\ \bibnamefont {Zhang}}, \ and\
  \bibinfo {author} {\bibfnamefont {Y~D.}\ \bibnamefont {Chong}},\ }\bibfield
  {title} {\enquote {\bibinfo {title} {Circuit implementation of a
  four-dimensional topological insulator},}\ }\href {\doibase
  10.1038/s41467-020-15940-3} {\bibfield  {journal} {\bibinfo  {journal}
  {Nature Communications}\ }\textbf {\bibinfo {volume} {11}},\ \bibinfo {pages}
  {2356} (\bibinfo {year} {2020})}\BibitemShut {NoStop}%
\bibitem [{\citenamefont {Olekhno}\ \emph {et~al.}(2020)\citenamefont
  {Olekhno}, \citenamefont {Kretov}, \citenamefont {Stepanenko}, \citenamefont
  {Ivanova}, \citenamefont {Yaroshenko}, \citenamefont {Puhtina}, \citenamefont
  {Filonov}, \citenamefont {Cappello}, \citenamefont {Matekovits},\ and\
  \citenamefont {Gorlach}}]{2020_Olekhno}%
  \BibitemOpen
  \bibfield  {author} {\bibinfo {author} {\bibfnamefont {Nikita~A.}\
  \bibnamefont {Olekhno}}, \bibinfo {author} {\bibfnamefont {Egor~I.}\
  \bibnamefont {Kretov}}, \bibinfo {author} {\bibfnamefont {Andrei~A.}\
  \bibnamefont {Stepanenko}}, \bibinfo {author} {\bibfnamefont {Polina~A.}\
  \bibnamefont {Ivanova}}, \bibinfo {author} {\bibfnamefont {Vitaly~V.}\
  \bibnamefont {Yaroshenko}}, \bibinfo {author} {\bibfnamefont {Ekaterina~M.}\
  \bibnamefont {Puhtina}}, \bibinfo {author} {\bibfnamefont {Dmitry~S.}\
  \bibnamefont {Filonov}}, \bibinfo {author} {\bibfnamefont {Barbara}\
  \bibnamefont {Cappello}}, \bibinfo {author} {\bibfnamefont {Ladislau}\
  \bibnamefont {Matekovits}}, \ and\ \bibinfo {author} {\bibfnamefont
  {Maxim~A.}\ \bibnamefont {Gorlach}},\ }\bibfield  {title} {\enquote {\bibinfo
  {title} {Topological edge states of interacting photon pairs emulated in a
  topolectrical circuit},}\ }\href {\doibase 10.1038/s41467-020-14994-7}
  {\bibfield  {journal} {\bibinfo  {journal} {Nature Communications}\ }\textbf
  {\bibinfo {volume} {11}},\ \bibinfo {pages} {1436} (\bibinfo {year}
  {2020})}\BibitemShut {NoStop}%
\bibitem [{\citenamefont {Hadad}\ \emph {et~al.}(2018)\citenamefont {Hadad},
  \citenamefont {Soric}, \citenamefont {Khanikaev},\ and\ \citenamefont
  {Al{\`{u}}}}]{2018_Hadad}%
  \BibitemOpen
  \bibfield  {author} {\bibinfo {author} {\bibfnamefont {Yakir}\ \bibnamefont
  {Hadad}}, \bibinfo {author} {\bibfnamefont {Jason~C.}\ \bibnamefont {Soric}},
  \bibinfo {author} {\bibfnamefont {Alexander~B.}\ \bibnamefont {Khanikaev}}, \
  and\ \bibinfo {author} {\bibfnamefont {Andrea}\ \bibnamefont {Al{\`{u}}}},\
  }\bibfield  {title} {\enquote {\bibinfo {title} {Self-induced topological
  protection in nonlinear circuit arrays},}\ }\href {\doibase
  10.1038/s41928-018-0042-z} {\bibfield  {journal} {\bibinfo  {journal} {Nature
  Electronics}\ }\textbf {\bibinfo {volume} {1}},\ \bibinfo {pages} {178--182}
  (\bibinfo {year} {2018})}\BibitemShut {NoStop}%
\bibitem [{\citenamefont {Bao}\ \emph {et~al.}(2019)\citenamefont {Bao},
  \citenamefont {Zou}, \citenamefont {Zhang}, \citenamefont {He}, \citenamefont
  {Sun},\ and\ \citenamefont {Zhang}}]{2019_Bao}%
  \BibitemOpen
  \bibfield  {author} {\bibinfo {author} {\bibfnamefont {Jiacheng}\
  \bibnamefont {Bao}}, \bibinfo {author} {\bibfnamefont {Deyuan}\ \bibnamefont
  {Zou}}, \bibinfo {author} {\bibfnamefont {Weixuan}\ \bibnamefont {Zhang}},
  \bibinfo {author} {\bibfnamefont {Wenjing}\ \bibnamefont {He}}, \bibinfo
  {author} {\bibfnamefont {Houjun}\ \bibnamefont {Sun}}, \ and\ \bibinfo
  {author} {\bibfnamefont {Xiangdong}\ \bibnamefont {Zhang}},\ }\bibfield
  {title} {\enquote {\bibinfo {title} {Topoelectrical circuit octupole
  insulator with topologically protected corner states},}\ }\href {\doibase
  10.1103/physrevb.100.201406} {\bibfield  {journal} {\bibinfo  {journal}
  {Physical Review B}\ }\textbf {\bibinfo {volume} {100}},\ \bibinfo {pages}
  {201406(R)} (\bibinfo {year} {2019})}\BibitemShut {NoStop}%
\bibitem [{\citenamefont {Liu}\ \emph {et~al.}(2020)\citenamefont {Liu},
  \citenamefont {Ma}, \citenamefont {Zhang}, \citenamefont {Zhang},
  \citenamefont {Yang}, \citenamefont {You}, \citenamefont {Gao}, \citenamefont
  {Xiang}, \citenamefont {Cui},\ and\ \citenamefont {Zhang}}]{2020_Liu}%
  \BibitemOpen
  \bibfield  {author} {\bibinfo {author} {\bibfnamefont {Shuo}\ \bibnamefont
  {Liu}}, \bibinfo {author} {\bibfnamefont {Shaojie}\ \bibnamefont {Ma}},
  \bibinfo {author} {\bibfnamefont {Qian}\ \bibnamefont {Zhang}}, \bibinfo
  {author} {\bibfnamefont {Lei}\ \bibnamefont {Zhang}}, \bibinfo {author}
  {\bibfnamefont {Cheng}\ \bibnamefont {Yang}}, \bibinfo {author}
  {\bibfnamefont {Oubo}\ \bibnamefont {You}}, \bibinfo {author} {\bibfnamefont
  {Wenlong}\ \bibnamefont {Gao}}, \bibinfo {author} {\bibfnamefont {Yuanjiang}\
  \bibnamefont {Xiang}}, \bibinfo {author} {\bibfnamefont {Tie~Jun}\
  \bibnamefont {Cui}}, \ and\ \bibinfo {author} {\bibfnamefont {Shuang}\
  \bibnamefont {Zhang}},\ }\bibfield  {title} {\enquote {\bibinfo {title}
  {Octupole corner state in a three-dimensional topological circuit},}\ }\href
  {\doibase 10.1038/s41377-020-00381-w} {\bibfield  {journal} {\bibinfo
  {journal} {Light: Science {\&} Applications}\ }\textbf {\bibinfo {volume}
  {9}},\ \bibinfo {pages} {145} (\bibinfo {year} {2020})}\BibitemShut {NoStop}%
\bibitem [{\citenamefont {Ningyuan}\ \emph {et~al.}(2015)\citenamefont
  {Ningyuan}, \citenamefont {Owens}, \citenamefont {Sommer}, \citenamefont
  {Schuster},\ and\ \citenamefont {Simon}}]{2015_Ningyuan}%
  \BibitemOpen
  \bibfield  {author} {\bibinfo {author} {\bibfnamefont {Jia}\ \bibnamefont
  {Ningyuan}}, \bibinfo {author} {\bibfnamefont {Clai}\ \bibnamefont {Owens}},
  \bibinfo {author} {\bibfnamefont {Ariel}\ \bibnamefont {Sommer}}, \bibinfo
  {author} {\bibfnamefont {David}\ \bibnamefont {Schuster}}, \ and\ \bibinfo
  {author} {\bibfnamefont {Jonathan}\ \bibnamefont {Simon}},\ }\bibfield
  {title} {\enquote {\bibinfo {title} {Time- and site-resolved dynamics in a
  topological circuit},}\ }\href {\doibase 10.1103/physrevx.5.021031}
  {\bibfield  {journal} {\bibinfo  {journal} {Physical Review X}\ }\textbf
  {\bibinfo {volume} {5}},\ \bibinfo {pages} {021031} (\bibinfo {year}
  {2015})}\BibitemShut {NoStop}%
\bibitem [{\citenamefont {Liu}\ \emph {et~al.}(2019)\citenamefont {Liu},
  \citenamefont {Gao}, \citenamefont {Zhang}, \citenamefont {Ma}, \citenamefont
  {Zhang}, \citenamefont {Liu}, \citenamefont {Xiang}, \citenamefont {Cui},\
  and\ \citenamefont {Zhang}}]{2019_Liu}%
  \BibitemOpen
  \bibfield  {author} {\bibinfo {author} {\bibfnamefont {Shuo}\ \bibnamefont
  {Liu}}, \bibinfo {author} {\bibfnamefont {Wenlong}\ \bibnamefont {Gao}},
  \bibinfo {author} {\bibfnamefont {Qian}\ \bibnamefont {Zhang}}, \bibinfo
  {author} {\bibfnamefont {Shaojie}\ \bibnamefont {Ma}}, \bibinfo {author}
  {\bibfnamefont {Lei}\ \bibnamefont {Zhang}}, \bibinfo {author} {\bibfnamefont
  {Changxu}\ \bibnamefont {Liu}}, \bibinfo {author} {\bibfnamefont
  {Yuan~Jiang}\ \bibnamefont {Xiang}}, \bibinfo {author} {\bibfnamefont
  {Tie~Jun}\ \bibnamefont {Cui}}, \ and\ \bibinfo {author} {\bibfnamefont
  {Shuang}\ \bibnamefont {Zhang}},\ }\bibfield  {title} {\enquote {\bibinfo
  {title} {Topologically protected edge state in two-dimensional
  {Su-Schrieffer-Heeger} circuit},}\ }\href {\doibase 10.1155/2019/8609875}
  {\bibfield  {journal} {\bibinfo  {journal} {Research}\ }\textbf {\bibinfo
  {volume} {2019}},\ \bibinfo {pages} {1--8} (\bibinfo {year}
  {2019})}\BibitemShut {NoStop}%
\end{thebibliography}%


\end{document}


\maketitle

\newpage

\tableofcontents


\newpage
\section{Supplementary Note 1 -- General properties of the extended Su-Schrieffer-Heeger model}
\label{sec:Theory_General}

\begin{center}
\begin{figure}
\begin{minipage}{\textwidth}
    \centerline{\includegraphics[width=12cm]{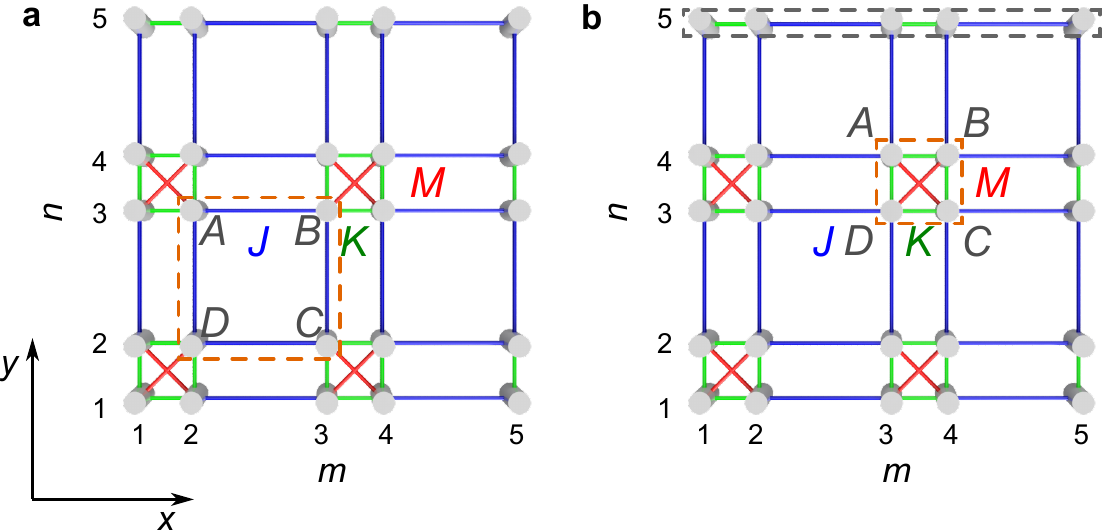}}
    \caption{The geometry of the considered tight-binding model.  Positive tunneling couplings $J$, $K$ and $M$ are shown by blue, green, and red solid lines, respectively. The dashed rectangles demonstrate two possible choices of the the unit cell. Panel \textbf{a} corresponds to the topological case (nonzero topological invariant), whereas panel \textbf{b} corresponds to the trivial case. Labels $A$, $B$, $C$, and $D$ mark the sites of the unit cell for the respective unit cell choices. Grey dashed line in panel \textbf{b} denotes the alignment of open boundary for the edge state dispersion calculation. \label{fig:S1_Model}}
\end{minipage}
\end{figure}
\end{center}

We consider an extended Su-Schrieffer-Heeger model with couplings $J>0$, $K>J$ and additional tunneling links $M>0$, as introduced in the main text of the article, Fig.~\ref{fig:S1_Model}a.

If the choice of the unit cell is consistent with Fig.~\ref{fig:S1_Model}b, the system is described by the following Bloch Hamiltonian:
\begin{equation}
    \hat{H}(k_x,k_y) = -\left(
    \begin{tabular}{c c c c}
    0 & $K + J{\rm e}^{-ik_{\rm x}}$ & $M$ & $K + J{\rm e}^{ik_{y}}$ \\
    $K + J{\rm e}^{ik_{\rm x}}$ & $0$ & $K + J{\rm e}^{ik_{\rm y}}$ & $M$ \\
    $M$ & $K + J{\rm e}^{-ik_{\rm y}}$ & $0$ & $K + J{\rm e}^{ik_{\rm x}}$ \\
    $K + J{\rm e}^{-ik_{\rm y}}$ & $M$ & $K + J{\rm e}^{-ik_{\rm x}}$ & $0$
    \end{tabular}
    \right),
\end{equation}
where $k_{\rm x}$ and $k_{\rm y}$ are the components of the wave vector along $x$- and $y$-axes of Fig.~\ref{fig:S1_Model}, and  columns of the Hamiltonian matrix from left to right correspond to the unit cell sites with indices $A$, $B$, $C$, and $D$, respectively.

\begin{center}
\begin{figure}
\begin{minipage}{\textwidth}
    \centerline{\includegraphics[width=10cm]{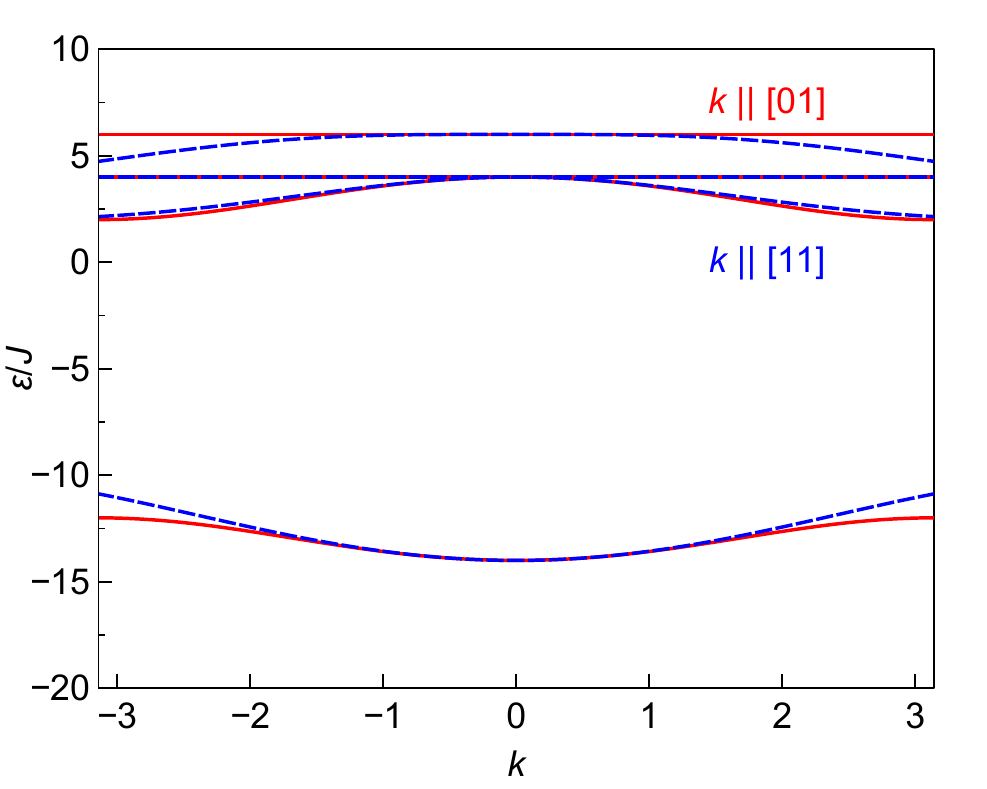}}
    \caption{Dispersion of four bulk bands in the considered model for wave vectors in directions $k||[01]$ (red solid line) and $k||[11]$ (blue dashed line). \label{fig:S1_1_Dispersion}}
\end{minipage}
\end{figure}
\end{center}

The energies of the eigenstates can be found as the solution to the secular equation ${\rm det}|\hat{H}(k_{x},k_{y}) -\varepsilon\hat{I}|=0$, where $\varepsilon$ is the energy of the eigenmode, and $\hat{I}$ is a $4 \times 4$ identity matrix. The resulting characteristic polynomial has the following form:
\begin{equation}
    A_{4}\varepsilon^{4} + A_{3}\varepsilon^{3} + A_{2}\varepsilon^{2} + A_{1}\varepsilon + A_{0} = 0,
    \label{eq:Dispersion}
\end{equation}
with coefficients
\begin{align*}
    A_{0} &= 4J^{2}K^{2} - 4K^{2}M^{2} + M^{4} -4JKM^{2}{\rm cos}(k_{\rm x}) + 2J^{2}(K-M)(K+M){\rm cos}(2k_{\rm x}) -\\
    &- 4JKM^{2}{\rm cos}(k_{\rm y}) - 8J^{2}K^{2}{\rm cos}(k_{\rm x}){\rm cos}(k_{\rm y}) + 2J^{2}(K-M)(K+M){\rm cos}(2k_{\rm y})\:,\\
    A_{1} &= 8K^{2}M + 8JKM{\rm cos}(k_{\rm x}) + 8JKM{\rm cos}(k_{\rm y}) + 8J^{2}M{\rm cos}(k_{\rm x}){\rm cos}(k_{\rm y})\:,\\
    A_{2} &= -4J^{2} - 4K^{2} - 2M^{2} - 4JK{\rm cos}(k_{\rm x}) - 4JK{\rm cos}(k_{\rm y})\:,\\
    A_{3} &= 0\:,\\
    A_{4} &= 1\:.
\end{align*}
In the general case, the roots of this characteristic polynomial are the functions of the wave vector components $k_{\rm x}$ and $k_{\rm y}$. Finding these solutions allows one to study the dispersion of the bulk bands. Besides, for the specific high-symmetry points, the obtained polynomial can be rewritten in the following simplified forms:
\begin{align}
    \Gamma~(k_{\rm x}=0,k_{\rm y}=0):& \quad (\varepsilon + M)^{2}(\varepsilon + 2J + 2K - M)(\varepsilon - 2J - 2K - M) = 0\:,\\
    {\rm X}~(k_{\rm x}=\pi,k_{\rm y}=0):& \quad (\varepsilon + 2J + M)(\varepsilon - 2J + M)(\varepsilon + 2K - M)(\varepsilon - 2K - M) = 0\:,\\
    {\rm M}~(k_{\rm x}=\pi,k_{\rm y}=\pi):& \quad (\varepsilon + M)^{2}(\varepsilon + 2J - 2K - M)(\varepsilon - 2J + 2K - M) = 0\:.
\end{align}
The resulting dispersion curves for bulk modes with wave vector directions $k||[01]$ (along the $x$-axis in Fig.~\ref{fig:S1_Model}) and $k||[11]$ (along the diagonal $x=y$) are shown in Fig.~\ref{fig:S1_1_Dispersion}. The dispersion diagram calculated for  $\Gamma-X-M-\Gamma$ trajectory in k-space discussed in the main text is also obtained by solving Eq.~(\ref{eq:Dispersion}) numerically.

To describe the dispersion of the edge states, we consider the semi-infinite geometry. This yields the linear system of equations
\begin{align*}
    -\varepsilon a + (J + K{\rm e}^{-ik_{\rm x}})b + MC + (J + K{\rm e}^{ik_{\rm y}})d &= 0,\\
    (J + K{\rm e}^{ik_{\rm x}})a - \varepsilon b + (J + K{\rm e}^{ik_{\rm y}})c + Md &= 0,\\
    Ma + (K + J{\rm e}^{-ik_{\rm y}})b - \varepsilon c + (K + J{\rm e}^{ik_{\rm x}})d &= 0,\\
    (K + J{\rm e}^{-ik_{\rm y}})a + Mb + (K + J{\rm e}^{-ik_{\rm x}})c - \varepsilon d &= 0,
    \label{eq:Edge_states}
\end{align*}
where $a$, $b$, $c$ and $d$ are the amplitudes of the wavefunction at the respective sites $A$, $B$, $C$, and $D$ of the unit cell shown in Fig.~\ref{fig:S1_Model}b, and $\varepsilon$ is the edge state energy. The above system should be supplemented by the following boundary conditions corresponding to the open boundary aligned along the $x$-axis, as shown in Fig.~\ref{fig:S1_Model}b:
\begin{align*}
    -\varepsilon a + (J + K{\rm e}^{-ik_{\rm x}})b + Jd &= 0,\\
    (J + K{\rm e}^{ik_{\rm x}})a - \varepsilon b + Jc &= 0.
\end{align*}
Then, the system reduces to the following form:
\begin{align*}
	-\varepsilon a + (J + K{\rm e}^{-ik_{\rm x}})b &= 0,\\
	(J + K{\rm e}^{ik_{\rm x}})a - \varepsilon b &= 0,\\
	c &= 0,\\
	d &= 0,
\end{align*}
i.e., the field amplitudes in $c$ and $d$ sublattices vanish. As a result, the energies of edge states read
\begin{equation}
    \varepsilon^{2} = J^{2} + K^{2} + 2JK{\rm cos}(k_{\rm x}),
    \label{eq:Edge_state_energy}
\end{equation}
which coincides with the dispersion of the bulk states in the one-dimensional (1D) Su-Schrieffer-Heeger model. The mode profiles of the edge states also coincide with those of the bulk states in 1D model.


Finally, our system supports a corner state localized at the corner with the weak links $J$. However, the analytical treatment of the corner state appears to be quite cumbersome and therefore we explored it by the exact diagonalization of the finite system Hamiltonian. Our calculations reveal that the energy of the corner state is quite close to zero, but deviates from the exact zero. At the same time, field profile of the corner mode features a behavior quite similar to the canonical Su-Schrieffer-Heeger model.

\newpage
\section{Supplementary Note 2 -- Topological invariant calculation}
\label{sec:Theory_Invariant}

To examine higher-order topology in the considered system with $D_4$ crystalline symmetry, we apply the technique\cite{2019_Benalcazar} evaluating the relevant topological invariants from the eigenvalues of $C_4$ rotation operator. Rotation by the angle $\pi/2$ for a four-site unit cell is defined by the operator:
\renewcommand{\arraystretch}{0.5}
\begin{equation}
    \hat{R}_{4} = 
    \begin{pmatrix}
        0 \ 1 \  0 \ 0\\ 0 \ 0 \ 1 \ 0 \\ 0 \ 0 \ 0 \ 1 \\ 1 \ 0 \ 0 \ 0
    \end{pmatrix},
\end{equation}
with columns of the operator matrix corresponding to the unit cell sites with indices $A$, $B$, $C$, and $D$ in Fig.~\ref{fig:S1_Model} from left to right, respectively. Such an operator has the eigenvalues $e^{2\pi i (p-1)/4}$ for $p=1,2,3,4$.
\renewcommand{\arraystretch}{1}

Due to the $D_4$ symmetry of the structure, there is a set of high-symmetry points $(k_x,k_y)$ in the first Brillouin zone such that $\hat{H}(k_x,k_y)\,\hat{R}_4=\hat{R}_4\,\hat{H}(k_x,k_y)$. Hence, the Hamiltonian and the rotation operator share the same set of eigenfunctions in such points, and the eigenstates can be labelled by the respective eigenvalues of the rotation operator, or, more simply, by the indices $p=1,2,3,4$. Counting the number of eigenstates with a specific value of $p$ below (or above) the chosen bandgap for the different high-symmetry points, one  determines the associated set of topological invariants\cite{2019_Benalcazar}. In such calculation, the labeling of eigenstates in the $\Gamma$ point in the center of Brillouin zone is taken as a reference:
\begin{equation}
    \chi^{(4)}\equiv 
    \begin{pmatrix}
        \#X_1^{(2)}-\#\Gamma_1^{(2)}\\
        \#M_1^{(4)}-\#\Gamma_1^{(4)}\\
        \#M_2^{(4)}-\#\Gamma_2^{(4)}
    \end{pmatrix},
    \label{eq:chi_definition}
\end{equation}
where the upper indices $(4)$ and $(2)$ correspond to the rotation by the angles $\pi/2$ and $\pi$ (double rotation by $\pi/2$), respectively, lower indices denote the values of parameter $p$ describing the considered rotation eigenvalues, while $\Gamma$, ${\rm M}$, and ${\rm X}$ letters denote the high-symmetry points of the first Brillouin zone with the coordinates $(k_{x},k_{y}) = (0,0)$, $(\pi, 0)$, and $(\pi,\pi)$, respectively. The symbol $\#$ in front of ${\rm X}$, ${\rm M}$, and $\Gamma$ denotes the number of energy bands below the bandgap with a given symmetry index $p$. 

To check the topological origin of the corner state localized at the weak link corner, we choose the unit cell accordingly, when amplitudes $J$ and $K$ are viewed as intra-cell and intercell couplings, respectively [Fig.~\ref{fig:S1_Model}a]. In such case, Bloch Hamiltonian takes the form
\begin{equation}
    \hat{H}(k_{x},k_{y})= 
        -\left(
    \begin{tabular}{c c c c}
    0 & $J+K e^{-i k_x}$ & $M e^{i k_y - i k_x}$  & $J+K e^{i  k_y}$ \\
        $J+K e^{i  k_x}$ & 0 & $J+K e^{i  k_y}$ & $M e^{i k_y + i k_x}$ \\
       $ M e^{-i k_y + i k_x}$ & $J+K e^{-i  k_y}$ & 0 & $J+K e^{i  k_x}$ \\
        $J+K e^{-i  k_y}$ & $M e^{-i k_y - i k_x}$ & $J+K e^{-i  k_x}$ & 0
    \end{tabular}
    \right),
    \label{eq:Bloch_Hamiltonian_Topological}
\end{equation}
and the commutation between $\hat{H}$ and $\hat{R}_4$ matrix in several high-symmetry points can be checked explicitly.


\subsection{Calculation for the topological case.}

In order to evaluate the topological invariant, one needs to calculate the eigenvalues and eigenstates of the Hamiltonian Eq.~\eqref{eq:Bloch_Hamiltonian_Topological} for the wave vectors $(k_x,k_y) = (0,0)$, $(\pi, 0)$, and $(\pi,\pi)$~\cite{2019_Benalcazar}.


\begin{table}
\begin{center}
\begin{tabular}{l l l c c}
\toprule
$\Gamma$-point & ($k_x=0$, $k_y=0$): & & \\
\toprule
 & $\centering{\varepsilon} $  & $\centering{\psi} $ & $\centering{p(R_{2})}$ & $\centering{p(R_{4})}$ \\
 \cmidrule(r){2-5}
 & $\varepsilon_1=-14$   &  $\psi_1=(1,1,1,1)$ & $1$ & $1$\\
 & $\varepsilon_2=4 $  &  $\psi_2=(i,-1,-i,1)$ & $2$ & $2$\\
 & $\varepsilon_3=4 $  &  $\psi_3=(-i,-1,i,1)$ & $2$ & $4$\\
 & $\varepsilon_4=6 $  &  $\psi_4=(-1,1,-1,1)$ & $1$ & $3$\\
 \toprule
${\rm X}$-point & ($k_x=\pi$, $k_y=0$): & & \\
 \toprule
  & $\centering{\varepsilon} $  & $\centering{\psi} $ & $\centering{p(R_{2})}$ &  \\
 \cmidrule(r){2-4}
 & $\varepsilon_1=-12$  &  $\psi_1=(1,-1,-1,1)$ & $2$ &\\
 & $\varepsilon_2=2$  &  $\psi_2=(1,1,1,1)$   & $2$ &\\
 & $\varepsilon_3=4$  &  $\psi_3=(-1,-1,1,1)$ & $1$ &\\
 & $\varepsilon_4=6$  &  $\psi_4=(-1,1,-1,1)$ & $1$ &\\
 \toprule
${\rm M}$-point & ($k_x=\pi$, $k_y=\pi$): & & \\
 \toprule
  & $\centering{\varepsilon} $  & $\centering{\psi} $ &  $\centering{p(R_{4})}$ & \\
 \cmidrule(r){2-5}
 & $\varepsilon_1=-10$   &  $\psi_1=(-1,1,-1,1)$ & $3$ &\\
  & $\varepsilon_4=2$  &  $\psi_2=(1,1,1,1)$   & $1$ &\\
 & $\varepsilon_3=4 $  &  $\psi_3=(i,-1,-i,1)$ & $2$ &\\
 & $\varepsilon_4=4 $  &  $\psi_4=(-i,-1,i,1)$ & $4$ &
\end{tabular}
\caption{\label{tab:Topological} Calculation of the topological invariant for topologically nontrivial case. $\varepsilon_{i}$ and $\psi_{i}$ are  eigenvalues and eigenfunctions of the Bloch Hamiltonian $H_0[k_x,k_y]$ Eq.(\ref{eq:Bloch_Hamiltonian_Topological}) for $i=1...4$, $p(R_2)$ and $p(R_4)$ are indices of eigenvalues of the  rotation operators $\hat{R}_2=\hat{R}_4^2$ and $\hat{R}_4$ by the angles $\pi$ and $\pi/2$ given by the expressions $\exp(i\pi(p(R_2)-1))$ and $\exp(i\pi(p(R_4)-1)/2)$, respectively.}
\end{center}
\end{table}

Using the table~\ref{tab:Topological} and checking the symmetry of the states below the bandgap around $\varepsilon\approx 0$, we recover:
\begin{align*}
    \#{\rm X}_{1}^{(2)}&=0, \quad \#\Gamma_{1}^{(2)} = 1,\\
    \#{\rm M}_{1}^{(4)}&=0, \quad \#\Gamma_{1}^{(4)} = 1,\\
    \#{\rm M}_{2}^{(4)}&=0, \quad \#\Gamma_{2}^{(4)} = 0.
\end{align*}
According to Eq.(\ref{eq:chi_definition}), this yields for vector $\chi^{(4)}$:
\begin{equation}
    \chi^{(4)}= 
    \begin{pmatrix}
        -1\\ -1 \\ 0
    \end{pmatrix},
\end{equation}
indicating the topological case. Nonzero invariant proves the existence of higher-order topological states with associated corner charge $Q_{\text{corner}}=\frac{1}{4}$ and the dipole polarization $\textbf{P}=(\frac{1}{2},\frac{1}{2})$ localized at the corner of the system formed by the weak couplings $J$.

\subsection{Calculation for the trivial case.}

In the opposite case, when the unit cell is chosen with the strong couplings inside [Fig.~\ref{fig:S1_Model}b], Bloch Hamiltonian of the system takes the form
\begin{equation}
    \hat{H}_{0}(k_x,k_y)= 
    -\left( \begin{tabular}{c c c c}
        $0$ & $K + J{\rm e}^{ik_{\rm x}}$ & $M$ & $K + J{\rm e}^{-ik_{\rm y}}$\\ 
        $K + J{\rm e}^{-ik_{\rm x}}$ & $0$ & $K + J{\rm e}^{-ik_{\rm y}}$ & $M$ \\ 
        $M$ & $K + J{\rm e}^{ik_{\rm y}}$ & $0$ & $K + J{\rm e}^{-ik_{\rm x}}$ \\
        $K + J{\rm e}^{ik_{\rm y}}$ & $M$ & $K + J{\rm e}^{ik_{\rm x}}$ & $0$
    \end{tabular}\right),
    \label{eq:Bloch_Hamiltonian_Trivial}
\end{equation}
Performing the same procedure as in the topological case, we obtain the set of eigenfunctions listed in the Table~\ref{tab:Trivial}.

\begin{table}
\begin{center}
\begin{tabular}{l l l c c}
\toprule
$\Gamma$-point & ($k_x=0$, $k_y=0$): & & \\
\toprule
 & $\centering{\varepsilon} $  & $\centering{\psi} $ & $\centering{p(R_{2})}$ & $\centering{p(R_{4})}$ \\
 \cmidrule(r){2-5}
 & $\varepsilon_1=-14$  &  $\psi_1=(1,1,1,1)$   & $1$ & $1$\\
 & $\varepsilon_2=4$  &  $\psi_2=(i,-1,-i,1)$ & $2$ & $2$\\
 & $\varepsilon_3=4$  &  $\psi_3=(-i,-1,i,1)$ & $2$ & $4$\\
 & $\varepsilon_4=6$  &  $\psi_4=(-1,1,-1,1)$ & $1$ & $3$\\
 \toprule
${\rm X}$-point & ($k_x=\pi$, $k_y=0$): & & \\
 \toprule
  & $\centering{\varepsilon} $  & $\centering{\psi} $ & $\centering{p(R_{2})}$ &  \\
 \cmidrule(r){2-4}
 & $\varepsilon_1=-12$  &  $\psi_1=(1,1,1,1)$   & $1$ &\\
 & $\varepsilon_2=2$  &  $\psi_2=(1,-1,-1,1)$ & $2$ &\\
 & $\varepsilon_3=4$  &  $\psi_3=(-1,1,-1,1)$ & $1$ &\\
 & $\varepsilon_4=6$  &  $\psi_4=(-1,-1,1,1)$ & $2$ &\\
 \toprule
${\rm M}$-point & ($k_x=\pi$, $k_y=\pi$): & & \\
 \toprule
  & $\centering{\varepsilon} $  & $\centering{\psi} $ &  $\centering{p(R_{4})}$ &\\
 \cmidrule(r){2-5}
 & $\varepsilon_1=-10$   &  $\psi_1=(1,1,1,1)$    & $1$ &\\
 & $\varepsilon_4=2$  &  $\psi_2=(-1,1,-1,1)$  & $3$ &\\
 & $\varepsilon_3=4$   &  $\psi_3=(i,-1,-i,1)$  & $2$ &\\
 & $\varepsilon_4=4$   &  $\psi_4=(-i,-1,i,1)$  & $4$ &
\end{tabular}
\caption{\label{tab:Trivial} Calculation of the topological invariant for topologically trivial case. $\varepsilon_{i}$ and $\psi_{i}$ are  eigenvalues and eigenfunctions of the Bloch Hamiltonian $H_0[k_x,k_y]$ Eq.(\ref{eq:Bloch_Hamiltonian_Trivial}) for $i=1...4$, $p(R_2)$ and $p(R_4)$ are indices of eigenvalues of the rotation operators $\hat{R}_2=\hat{R}_4^2$ and $\hat{R}_4$ by the angles $\pi$ and $\pi/2$ given by the expressions $\exp(i\pi(p(R_2)-1))$ and $\exp(i\pi(p(R_4)-1)/2)$, respectively.}
\end{center}
\end{table}

The symmetry of the eigenfunctions yields:
\begin{align*}
    \#{\rm X}_{1}^{(2)}&=1, \quad \#\Gamma_{1}^{(2)} = 1,\\
    \#{\rm M}_{1}^{(4)}&=1, \quad \#\Gamma_{1}^{(4)} = 1,\\
    \#{\rm M}_{2}^{(4)}&=0, \quad \#\Gamma_{2}^{(4)} = 0,
\end{align*}
and the vector $\chi^{(4)}$ equals
\begin{equation}
    \chi^{(4)}= 
    \begin{pmatrix}
        0\\ 0 \\ 0
    \end{pmatrix}.
\end{equation}
This trivial result indicates that higher-order topological states are absent at the corners of the system formed by the strong links $K$ which agrees with our tight-binding calculation.

\newpage
\section{Supplementary Note 3 -- Anti-symmetric bound states in the continuum in the extended Su-Schrieffer-Heeger model}
\label{sec:Theory_BIC}

\begin{center}
\begin{figure}
\begin{minipage}{\textwidth}
    \centerline{\includegraphics[width=16cm]{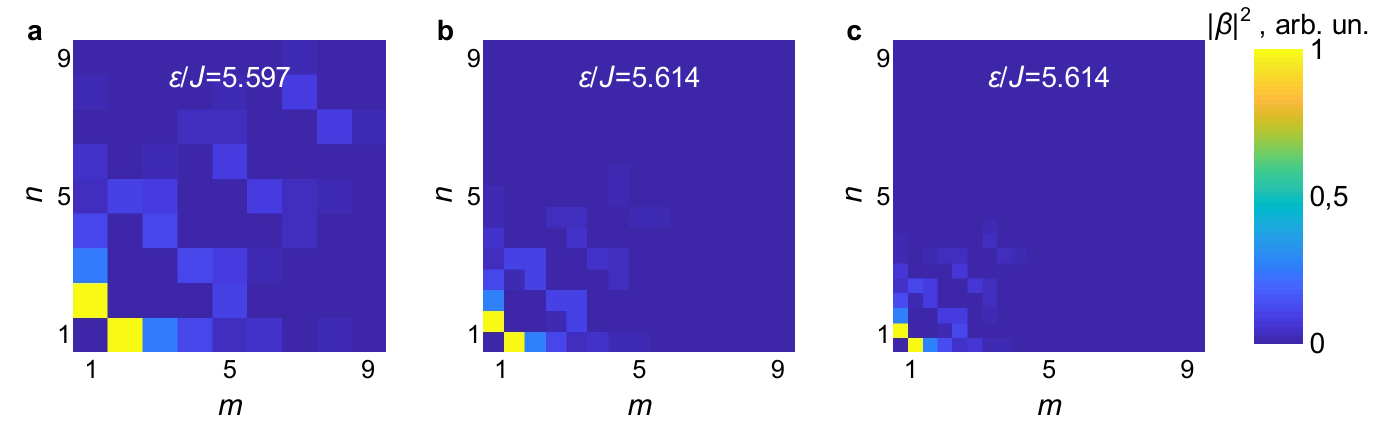}}
    \caption{Eigenmodes corresponding to the BIC Type II corner states in the extended SSH model with $K=4$ and $M=5.085$ having different number of sites: $9 \times 9$ (\textbf{a}), $15 \times 15$ (\textbf{b}), and $21 \times 21$ (\textbf{c}).\label{fig:S2_BIC}}
\end{minipage}
\end{figure}
\end{center}

In this Note, we examine the profiles of eigenmodes denoted in the article main text as BIC Type II corner states. In a relatively small system of $9 \times 9$ sites, such state can resemble a bulk mode (see Fig.~2 in the main text). However, when the size of the system is increased, the difference of this state from bulk modes becomes more evident. Specifically, Fig.~\ref{fig:S2_BIC}a-c shows that the corresponding state does not change its geometry and remains localized with the increase of the system size, thus demonstrating the bound nature. One can also observe a small shift in the energy of the associated state for the system size $9 \times 9$ (Fig.~\ref{fig:S2_BIC}a), while for the sufficiently large systems the energy of the state becomes insensitive to the number of sites (Fig.~\ref{fig:S2_BIC}b,c).

The emergence of bound state in the continuum highlights additional features of the proposed model aside from its topological origin.

\newpage
\section{Supplementary Note 4 -- Construction of equivalent electric circuit}
\label{sec:Circuit_Mapping}

\begin{center}
\begin{figure}
\begin{minipage}{\textwidth}
    \centerline{\includegraphics[width=16cm]{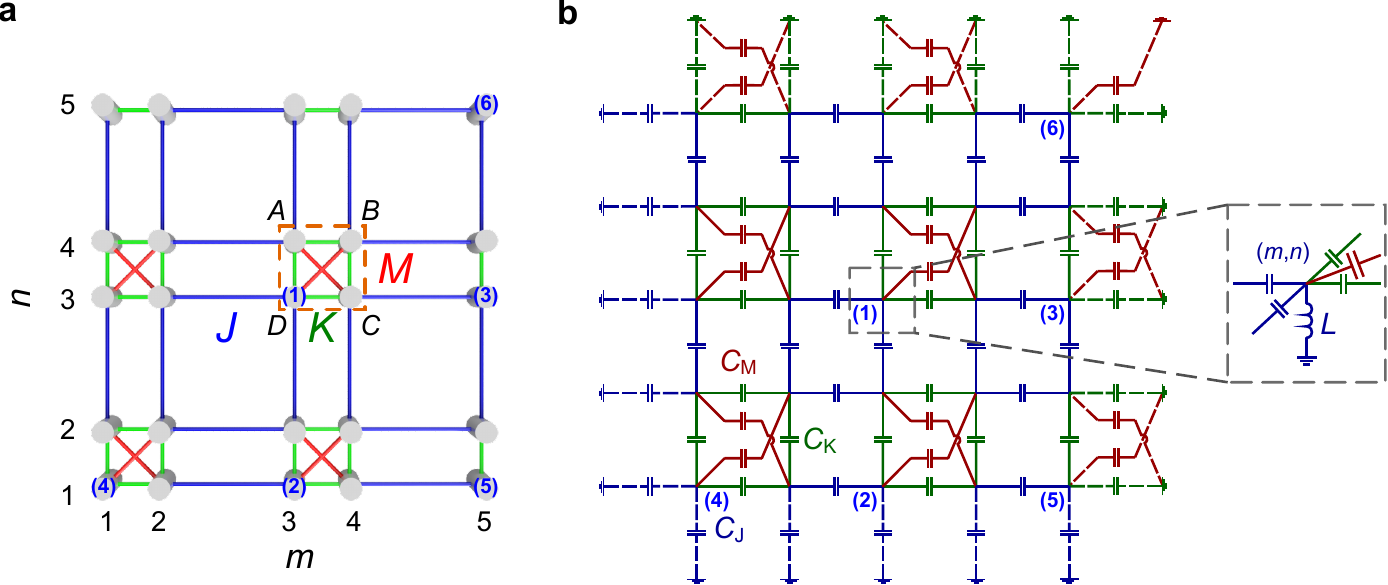}}
    \caption{Mapping between the extended SSH model (\textbf{a}) and the equivalent electric circuit (\textbf{b}) for the system with $N=5$ sites at the edge taken as an example. Indices in brackets mark the full set of sites for which a mapping between the tight-binding model and Kirchhoff's rules should be established. Labels A, B, C and D in panel \textbf{a} denote four sites of the unit cell.\label{fig:S3_Circuit}}
\end{minipage}
\end{figure}
\end{center}

We start with writing a full system of tight-binding equations for the finite extended SSH model of the size $N \times N$ sites with odd number of sites at the edge $N$, Fig.~\ref{fig:S3_Circuit}a. Due to the $D_{4}$ symmetry of the unit cell, such a system has six types of different sites for which the tight-binding equations should be formulated. Such sites are marked in Fig.~\ref{fig:S3_Circuit}a. The tight-binding equations for the sites (1-6) have the following form:
\begin{align}
    \label{eq:Mapping_TB_1}
    (1):& \quad -J(\beta_{m,n-1} + \beta_{m-1,n}) - K(\beta_{m,n+1} +\beta_{m+1,n}) - M\beta_{m+1,n+1} = \varepsilon\beta_{m,n}\\
    \label{eq:Mapping_TB_2}
    (2):& \quad -J\beta_{m-1,1} - K(\beta_{m+1,1}+\beta_{m,2}) - M\beta_{m+1,2} = \varepsilon\beta_{m,1}\\
    \label{eq:Mapping_TB_3}
    (3):& \quad -J(\beta_{N-1,m}+\beta_{N,m-1}) - K\beta_{N,m+1} = \varepsilon\beta_{N,m}\\
    \label{eq:Mapping_TB_4}
    (4):& \quad -K(\beta_{1,2}+\beta_{2,1}) - M\beta_{2,2} = \varepsilon\beta_{1,1}\\
    \label{eq:Mapping_TB_5}
    (5):& \quad -J\beta_{N-1,1} - K\beta_{N,2} = \varepsilon\beta_{N,1}\\
    \label{eq:Mapping_TB_6}
    (6):& \quad -J(\beta_{N-1,N} + \beta_{N,N-1}) = \varepsilon\beta_{N,N}
\end{align}
Equation (\ref{eq:Mapping_TB_1}) describes a bulk site of the extended SSH model. Due to the $D_{4}$ symmetry, the equations for the other sublattices of the unit cell can be obtained via rotation. Thus, it is sufficient to consider just one site of the unit cell in order to establish the mapping between the tight-binding model and electric circuit. Equations (\ref{eq:Mapping_TB_2}) and (\ref{eq:Mapping_TB_3}) describe edge sites of two different types, one with additional coupling $M$, site (2) in Fig.~\ref{fig:S3_Circuit}a, and another one without it, site (3) in Fig.~\ref{fig:S3_Circuit}a. As in the previous case of bulk sites, all other edge sites can be obtained either by a translation or rotations of the two mentioned. Finally, sites (4-6) represent three non-identical corners of the model.

Next, we consider the electric circuit depicted in Fig.~\ref{fig:S3_Circuit}b with the nodes connected by capacitors $C_{\rm J}$, $C_{\rm K}$, and $C_{\rm M}$, and grounded with inductors $L$. In the system bulk, the types and amount of bonds resemble those of the initial tight-binding model. However, the boundaries of the two models behave differently, and instead of open boundary conditions for the tight-binding model, the edge nodes of the circuit require additional grounding, as shown in Fig.~\ref{fig:S3_Circuit}b, in order to ensure the correct mapping. More precisely, the edge nodes should be grounded exactly with the elements that are lacking for the boundary elements.

The spectrum of the circuit and corresponding eigenmodes can be found from the linear system of Kirchhoff's rules describing currents and voltages in the circuit. Namely, one needs to consider the Kirchhoff's rule $\sum_{mn,m'n'~\in~node} I_{mn,m'n'} = 0$ stating that the sum of all currents $I_{mn,m'n'}$ flowing into an arbitrary node $(m,n)$ from its neighbors $(m',n')$ vanishes. The second Kirchhoff's rule $\sum_{mn,m'n'~\in~loop}U_{mn,m'n'}=0$ stating that the sum of voltage drops $U_{mn,m'n'}$ over the closed loop also vanishes, can be satisfied by introducing the on-site electric potentials $\varphi_{mn}$. Combined with Ohm's law $I_{mn,m'n'} = \sigma_{mn,m'n'}(\varphi_{m'n'} - \varphi_{mn})$ that relates the currents with potentials by means of the admittance $\sigma_{mn,m'n'}$ between the nodes $(m,n)$ and $(m',n')$, the linear system of Kirchhoff's current rules can be brought to the form 
\begin{align*}
    (1):& \quad -\sigma_{C_{\rm J}}(\varphi_{m,n-1} + \varphi_{m-1,n}) - \sigma_{C_{\rm K}}(\varphi_{m,n+1} +\varphi_{m+1,n}) -\\
    & \hspace{5cm} -\sigma_{C_{\rm M}}\varphi_{m+1,n+1} = (-2\sigma_{C_{\rm J}} - 2\sigma_{C_{\rm K}} - \sigma_{C_{\rm M}} - \sigma_{\rm L})\varphi_{m,n}\\
    (2):& \quad -\sigma_{C_{\rm J}}\varphi_{m-1,1} - \sigma_{C_{\rm K}}(\varphi_{m+1,1}+\varphi_{m,2}) - \sigma_{C_{\rm M}}\varphi_{m+1,2} = (-2\sigma_{C_{\rm J}} - 2\sigma_{C_{\rm K}} - \sigma_{C_{\rm M}} - \sigma_{\rm L})\varphi_{m,1}\\
    (3):& \quad -\sigma_{C_{\rm J}}(\varphi_{N-1,m}+\varphi_{N,m-1}) - \sigma_{C_{\rm K}}\varphi_{N,m+1} = (-2\sigma_{C_{\rm J}} - 2\sigma_{C_{\rm K}} - \sigma_{C_{\rm M}} - \sigma_{\rm L})\varphi_{N,m}\\
    (4):& \quad - \sigma_{C_{\rm K}}(\varphi_{1,2}+\varphi_{2,1}) - \sigma_{C_{\rm M}}\varphi_{2,2} = (-2\sigma_{C_{\rm J}} - 2\sigma_{C_{\rm K}} - \sigma_{C_{\rm M}} - \sigma_{\rm L})\varphi_{1,1}\\
    (5):& \quad -\sigma_{C_{\rm J}}\varphi_{N-1,1} - \sigma_{C_{\rm K}}\varphi_{N,2} = \varepsilon\varphi_{N,1}\\
    (6):& \quad -\sigma_{C_{\rm J}}(\varphi_{N-1,N} + \varphi_{N,N-1}) = (-2\sigma_{C_{\rm J}} - 2\sigma_{C_{\rm K}} - \sigma_{C_{\rm M}} - \sigma_{\rm L})\varphi_{N,N}
\end{align*}
with $\sigma_{C\rm J}=-i\omega C_{\rm J}$, $\sigma_{C\rm K}=-i\omega C_{\rm K}$, $\sigma_{C\rm M}=-i\omega C_{\rm M}$, and $\sigma_{L}=i/\omega L$ being the admittances of the bonds in the circuit. Note that we use $\varphi_{mn} \propto {\rm e}^{-i\omega t}$ notation for time-varying fields instead of the common radio-physics notation $\varphi_{mn} \propto {\rm e}^{i\omega t}$ for consistency with the Schr{\"o}dinger-type equation describing the tight-binding model. As seen from the obtained linear system, the coefficient before $\varphi_{mn}$ at the right hand side is the same for all types of nodes in the circuit, which results from the proper grounding of the edge nodes. Then, dividing the above system by $\sigma_{C_J}$, one finally obtains the set of equations Eq.~(\ref{eq:Mapping_TB_1}-\ref{eq:Mapping_TB_6}) with $J=1$, $K=C_{K}/C_{J}$, $M=C_{M}/C_{J}$, and the energy variable
\begin{equation*}
    \varepsilon = -2\Big(1 + \frac{C_{K}}{C_{J}}\Big) - \frac{C_{M}}{C_{J}} + \frac{1}{\omega^{2}LC_{J}} = \frac{f_{0}^2}{f^2} - 2(1 + K) - M\:,
\end{equation*}
defining the relation between the energies $\varepsilon$ of the eigenmodes in the tight-binding model and resonant frequencies $f$ of the equivalent circuit. Here, the characteristic frequency $f_0$ is defined as $f_0=1/(2\pi\,\sqrt{L\,C_J})$.

\newpage
\section{Supplementary Note 5 -- Experimental setup and measurement protocols}
\label{sec:Circuit_Experiments}

\subsection{Inductors preparation.}
The grounding inductors in the experimentally realized $9 \times 9$ circuit are Bourns RLB1314-220KL. These inductors serve as a main source of ohmic losses in our experimental setup, and also contribute to the diagonal disorder in the associated Hamiltonian. Besides, they can demonstrate frequency-dependent inductance and resistance due to the presence of a ferrite core and skin effect in the copper wire, changing their parameters upon a mechanical deformation. Thus, their careful preparation strongly affects circuit performance.

\begin{center}
\begin{figure}
\begin{minipage}{\textwidth}
    \centerline{\includegraphics{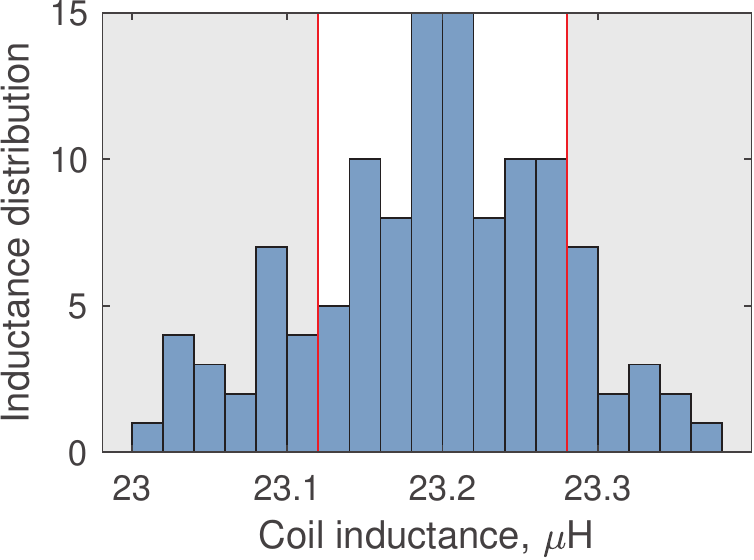}}
    \caption{A histogram showing the distribution of grounding coils inductance after the lead gluing and the first sorting procedure. The region shown with red solid lines and white background indicates $81$ inductors which were used in the experimental setup.\label{fig:S4_Inductors_Distribution}}
\end{minipage}
\end{figure}
\end{center}

The inductors with ferrite core used in our experimental setup have only around $20$ turns of copper wire, so the deformation of their leads during the circuit board assembly might potentially influence the values of inductance. To prevent this, each inductor was first stripped off the insulation sleeve which covers the ferrite rod and the wire. Next, the leads were fixed to the ferrite rod with the help of glue. Then, the inductors were sorted out with a Mastech MS5308 LCR-meter at the frequency of $10~{\rm kHz}$ characteristic of resonances in the circuit. Obtained inductance values were written on the stickers attached to the inductors with inductance values in the range $22.01 ... 23.37 \mu {\rm H}$.

After that, a statistical analysis of these inductors was conveyed. The distribution of the inductances is shown in Fig.~\ref{fig:S4_Inductors_Distribution}. As seen from the figure, the distribution is close to normal with mean inductance $23.19 \,\mu \rm{H}$ and variance $0.08 \,\mu \rm{H}$. Finally, $81$ inductors were chosen from this set, starting from the mean value. The associated range of inductances is shown in red in Fig.~\ref{fig:S4_Inductors_Distribution}.

\begin{center}
\begin{figure}
\begin{minipage}{\textwidth}
    \centerline{\includegraphics{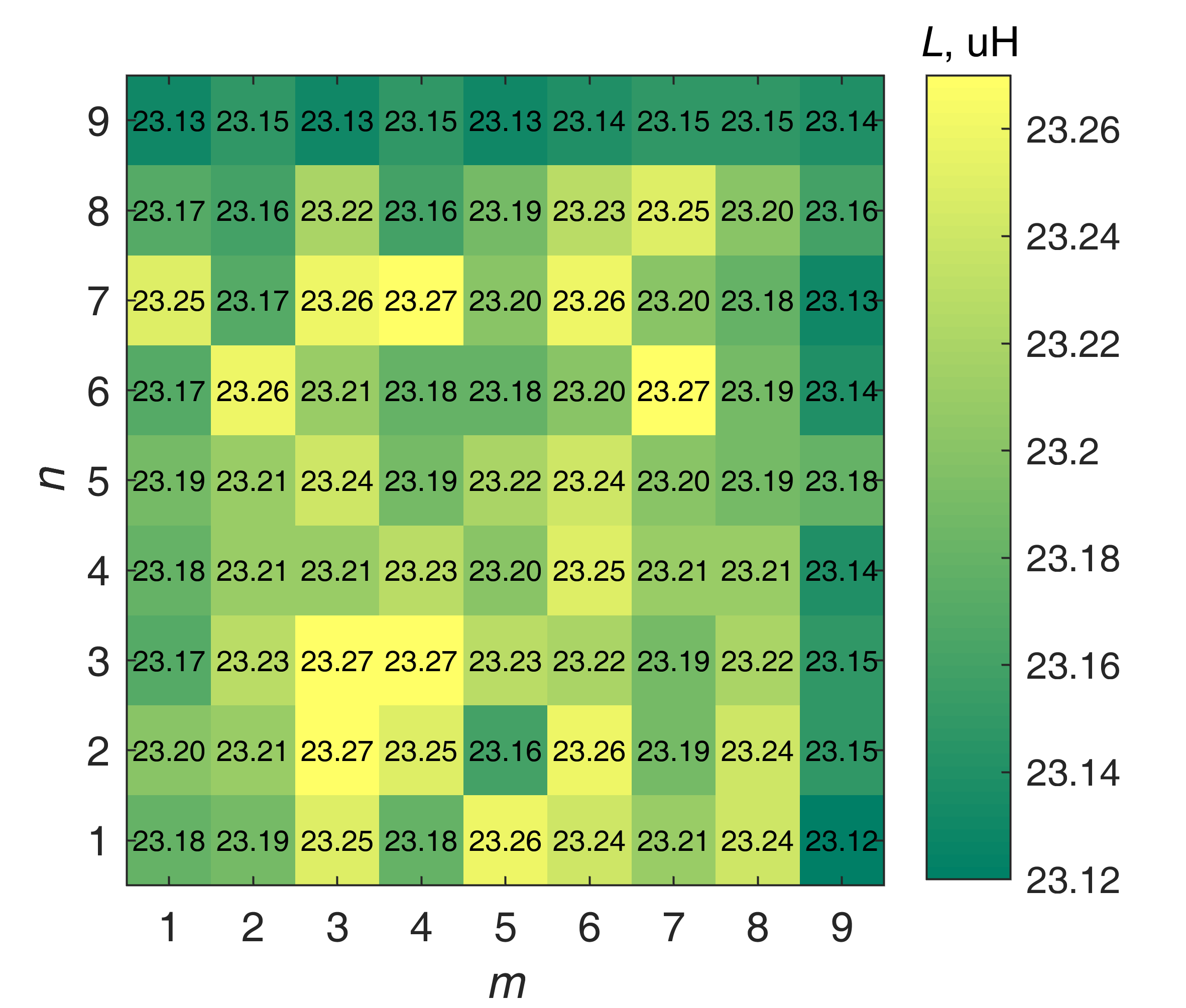}}
    \caption{Inductances of the grounding coils at each node of the experimentally realized topolectrical circuit. The color of the node visualizes the deviation from the minimal value of inductance (shown with green) to the largest one (shown with yellow). \label{fig:S5_Inductors_Map}}
\end{minipage}
\end{figure}
\end{center}

These inductors were then mounted on the PCB in a pseudo-random fashion. After soldering, the stickers were collected from the inductors to map the on-site inductance distribution depicted in Fig.~\ref{fig:S5_Inductors_Map}.

\subsection{Capacitors.}
In the experimentally realized circuit, we use Murata GRM32RR71H105KA01L as capacitors $C_J$ and Murata GRM31CR71H475KA12L as capacitors $C_{K}$ and $C_{M}$, both models being ceramic capacitors for surface mounting (surface-mount devices, SMD). In order to sort the capacitors, we used Mastech MS8910 and Mastech MS5308 LCR-meters. We have selected $82$ capacitors with values $C_{\rm J} = (931 \pm 3)\,{\rm nF}$, and $131$ capacitors with values $C_{\rm K} = C_{\rm M} = (4020 \pm 10)\,{\rm nF}$ yielding a tolerances of $\pm 0.32\%$ for capacitors $C_{\rm J}$, and $\pm 0.25\%$ for capacitors $C_{\rm K}$ $C_{\rm M}$, respectively. Selected capacitors were randomly distributed between the corresponding bonds of the circuit, without constructing a map as for inductors.

\subsection{Printed circuit board assembly.}

The printed circuit board was assembled using a $\text{Sn}_{97}\text{Cu}_{3}$ Pb-free solder and rosin gel flux. Soldering station temperature was set to $360^{\circ}$C for capacitors and inductors and to $380^{\circ}$C for MCX-type coaxial cable connectors. The soldering time was controlled not to exceed $4$ seconds per lead for inductors and connectors, and $4$ seconds per component for capacitors. After mounting all the components, the PCB was cleaned off the remaining flux using first warm water and soap, and after that $97\%$ isopropyl alcohol.

\subsection{Description of OSA103 Mini measurement equipment.}

OSA103 Mini is an open-source platform for electrical measurements at frequencies up to $100~{\rm MHz}$ which allows to perform the studies of frequency-dependent on-site response automatically, as well as to extract a full distribution of on-site potentials including relative phases. It consists of a direct digital synthesis (DDS) signal generator and an analog-to-digital converter (ADC). The DDS signal generator is based on a 12-bit R-2R digital-to-analog converter assembled with $1\%$ precision resistors connected to the general-purpose input/output pins of the FPGA. The ADC chip used is AD9288BST -- a 2 channel 8-bit 100~MSPS converter. The system is implemented on XILINX XC6SLX9-2TQG144C FPGA circuit, connected to a PC for control and data acquisition via UART-over-USB (using FT232RL chip).

The dynamic range in the frequency response analyzer mode is $90~{\rm dB}$. In the frequency analyzer mode, the device is calibrated to the $0~{\rm dB}$ level using an input-to-output loop before performing the actual measurements of the topological circuit frequency response. In our measurements, we used the driving signal level $-10~{\rm dBm}$.

\begin{center}
\begin{figure}
\begin{minipage}{\textwidth}
    \centerline{\includegraphics[width=8cm]{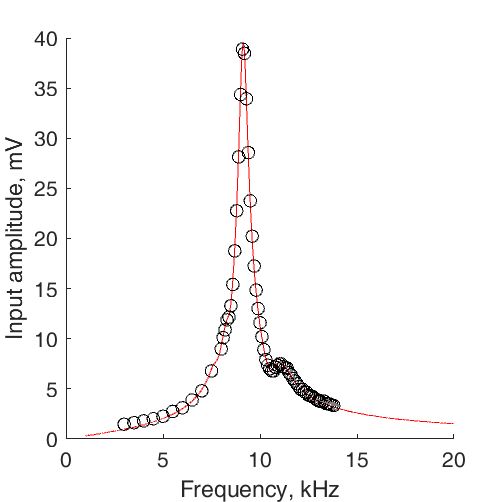}}
    \caption{Direct comparison of the frequency response at node $(9,9)$ of the experimentally realized topological circuit measured with OSA103 Mini (red solid line) and manually with Rohde\&Schwarz HMO 2022 oscilloscope and Keithley 3390 signal generator (black circles).\label{fig:S6_OSA_Verification}}
\end{minipage}
\end{figure}
\end{center}

In order to verify the results obtained with OSA103 Mini, we performed a direct check of the on-site voltage response at node $(9,9)$ of the experimental circuit, featuring a single resonance peak corresponding to the topological corner state, Fig.~\ref{fig:S6_OSA_Verification}. To do so, we first attached OSA103 Mini to the circuit using two MCX-type connectors at site $(9,9)$ for the input and output channels and performed automatic scan of the voltage between site $(9,9)$ and ground at $1000$ frequency points uniformly spanned between $20~{\rm Hz}$ and $20~{\rm kHz}$, and then performed the same measurement manually driving the circuit at given frequency with Keithley 3390 signal generator and measuring the voltage between site $(9,9)$ and ground with Rohde\&Schwarz HMO 2022 oscilloscope considering $100$ frequency points. In both cases, the driving voltage level was set to $120~{\rm mV}$. As seen from Fig.~\ref{fig:S6_OSA_Verification}, a perfect agreement between automatic and manual measurements is observed, which allowed us to perform the measurements of frequency-dependent voltage response at all sites of the circuit, as outlined in the main text.

\subsection{Measurement protocols.}
To characterize the properties of experimentally realized topological circuit, we measure the voltage between the given node of the circuit and ground when the harmonic signal at frequency $f$ is applied between the studied node and the ground. The peak-to-peak voltage of the driving signal is set to the constant value $U_{\rm ext}=120~{\rm mV}$ in the entire frequency range in the regime of $50~{\rm Ohm}$ series load for the output stage. Thus, the measured voltage response characterizes the full impedance of the circuit between the given node and ground. We perform such a measurement at every node of the circuit, obtaining $81$ voltage vs frequency curves each having $1000$ frequency points in the range from $20~{\rm Hz}$ to $20~{\rm kHz}$. This complete set of measurements allows us to extract full maps of on-site potential distributions at $1000$ frequency points, correspondingly, and thus reconstruct mode profiles of characteristic resonances shown in Fig.3 of the main text, as well as study their evolution with a change in the driving frequency.

\newpage
\section{Supplementary Note 6 -- Circuit simulations and disorder effects}
\label{sec:Circuit_Simulations}
To bridge the gap between theory and experiments and analyse the phenomena arising in electric circuit in more detail, we have performed numerical simulations of the equivalent circuit with Keysight Advanced Design System (ADS) software package. Three distinct cases have been considered: 
\begin{itemize}
    \item Idealized ordered circuit consisting of capacitors $C_{\rm{J}} = 1.00 \mu \rm{F}$, $C_{\rm{K}}=C_{\rm{M}}=4.00 \mu \rm{F}$ and grounding inductors $L=23.20 \mu \rm{H}$ without any fluctuations of element values
    \item Realistic circuit with randomly chosen values of elements within uniformly distributed $\pm 1\%$ disorder with $C_{\rm{J}} = 1.00 \pm 0.01 \mu \rm{F}$, $C_{\rm{K}}=C_{\rm{M}}=4.00 \pm 0.04 \mu \rm{F}$, $L=23.20 \pm 0.23 \mu \rm{H}$
    \item Strongly disordered circuit with randomly chosen values of elements within uniformly distributed $\pm 10\%$ disorder with $C_{\rm{J}} = 1.00 \pm 0.10 \mu \rm{F}$, $C_{\rm{K}}=C_{\rm{M}}=4.00 \pm 0.40 \mu \rm{F}$, $L=23.20 \pm 2.30 \mu \rm{H}$
\end{itemize}
In all three considered cases, ohmic losses have been taken into account by setting the $Q$-factor of inductors defined as the ratio of reactive and active impedances $Q_{\rm L}=2\pi f L/R_{\rm L}$ to be equal $Q_{\rm L}=43$ in the entire frequency region in accordance with experimental measurements indicating such a $Q$-factor at the frequency $f=10~{\rm kHz}$. Also, a ${\rm DC}$ series resistance $R_{\rm{DC}}=0.03~\rm{Ohm}$ have been added to all the grounding inductors $L$ according to the low-frequency measurements.

\begin{center}
\begin{figure}
\begin{minipage}{\textwidth}
    \centerline{\includegraphics[width=16cm]{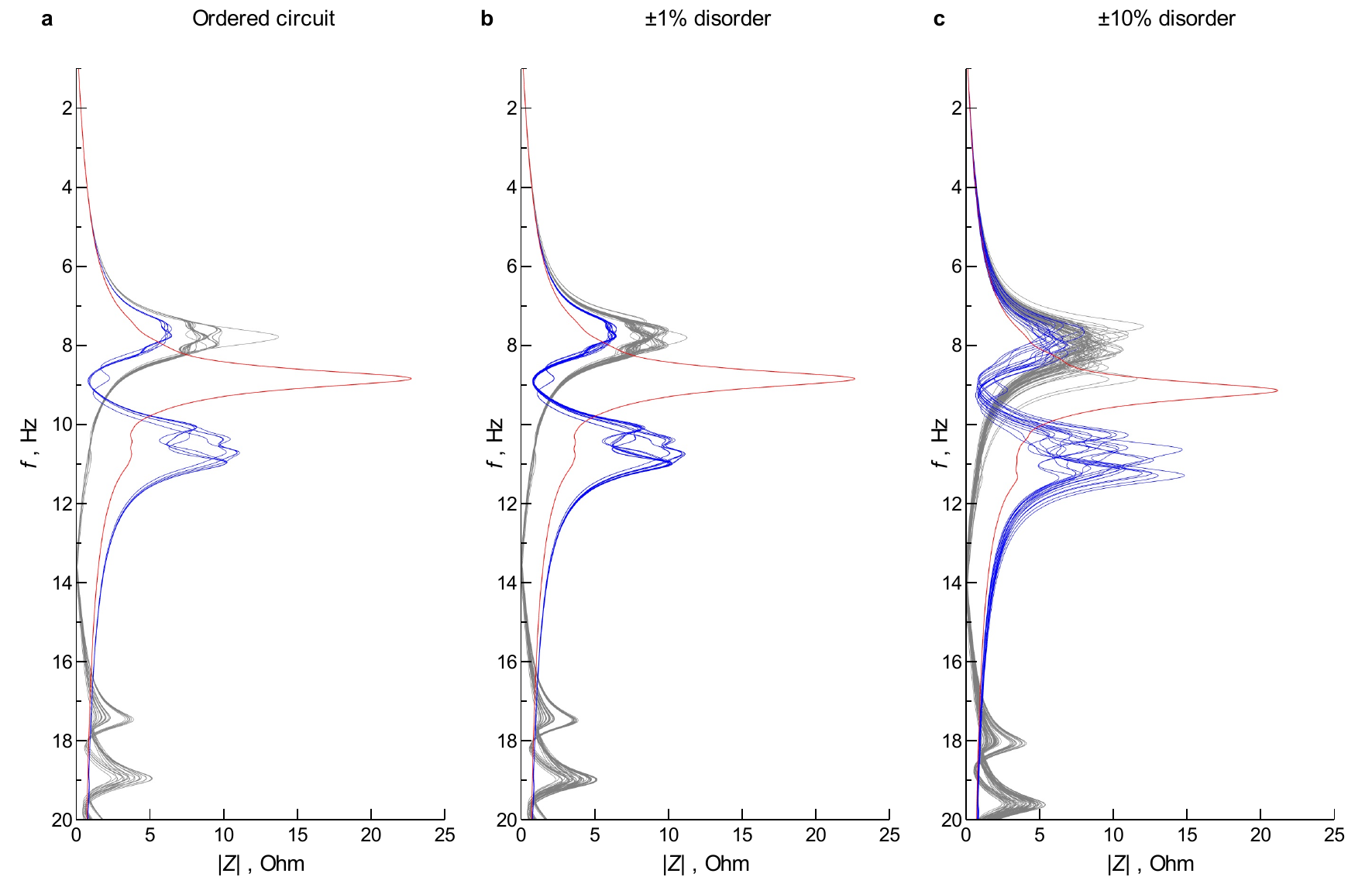}}
    \caption{Simulated spectra of the absolute value of impedance between the given node of the circuit and ground for the idealized circuit without disorder in element values (\textbf{a}), realistic circuit with $\pm 1\%$ disorder in element values (\textbf{b}) and strongly disordered circuit with $\pm 10\%$ disorder (\textbf{c}). Grey curves correspond to bulk and edge nodes with coordinates $1 \leq m$, $n \leq 8$, blue curves correspond to the edge nodes with $1 \leq 8$, $n=9$ and $1 \leq n \leq 8$, $m=9$ and red curves represent the corner node with coordinates $m=9$, $n=9$. \label{fig:S7_Spectra}}
\end{minipage}
\end{figure}
\end{center}

\begin{center}
\begin{figure}
\begin{minipage}{\textwidth}
    \centerline{\includegraphics[width=16cm]{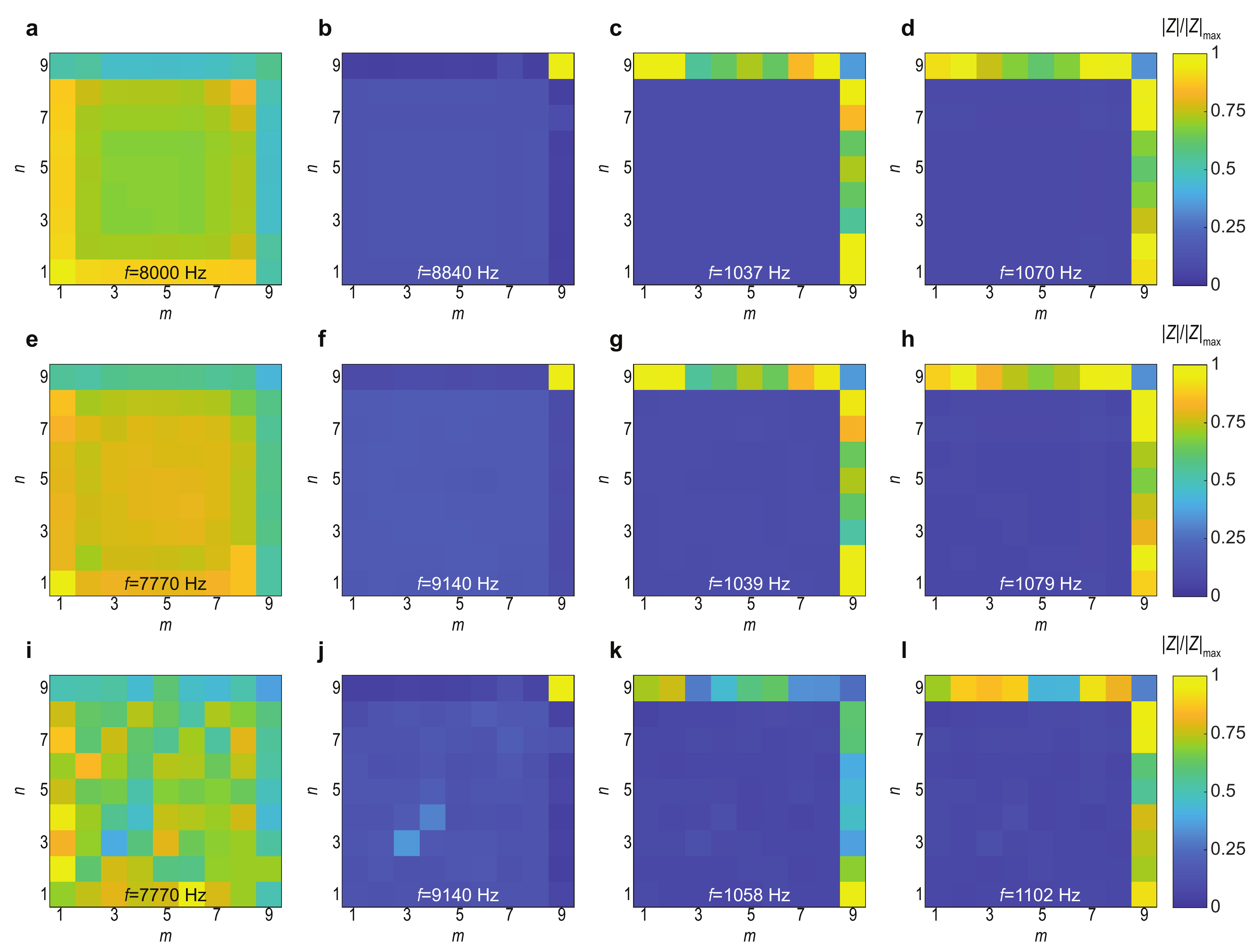}}
    \caption{Maps of the absolute value of on-site impedance at fixed driving frequencies $f$ in circuits with different degrees of disorder in capacitors $C_{\rm{J}}$, $C_{\rm{K}}$, $C_{\rm{M}}$ and grounding inductors $L$. \textbf{a-d}: the idealized circuit without disorder in element values ($C_{\rm{J}} = 1.00 \mu \rm{F}$, $C_{\rm{K}}=C_{\rm{M}}=4.00 \mu \rm{F}$, $L=23.20 \mu \rm{H}$ ). 
    \textbf{e-h}: the circuit with 1\% disorder ($C_{\rm{J}} = 1.00 \pm 0.01 \mu \rm{F}$, $C_{\rm{K}}=C_{\rm{M}}=4.00 \pm 0.04 \mu \rm{F}$, $L=23.20 \pm 0.23 \mu \rm{H}$). \textbf{i-l}: the circuit with 10\% disorder  ($C_{\rm{J}} = 1.00 \pm 0.10 \mu \rm{F}$, $C_{\rm{K}}=C_{\rm{M}}=4.00 \pm 0.40 \mu \rm{F}$, $L=23.20 \pm 2.30 \mu \rm{H}$). The modes represent bulk (\textbf{a, e, i}), edge (\textbf{c, d, g, h, k, i}) and corner (\textbf{b, f, j}) states. \label{fig:S8_Modes}}
\end{minipage}
\end{figure}
\end{center}

It is seen that the results of ordered circuit simulations demonstrate good agreement with tight-binding calculations. Indeed, impedance spectra at the nodes symmetrically located with respect to the diagonal $(m=n)$ coincide, as seen in Fig.~\ref{fig:S7_Spectra}a, and the modes of the circuit are symmetric in accordance with the circuit symmetry, Fig.~\ref{fig:S8_Modes}a-d. However, in contrast to theoretically obtained energy distributions, the spectrum in Fig.~\ref{fig:S7_Spectra}a features considerable {\it homogeneous} broadening of characteristic resonances associated with ohmic losses in inductors.

If some minimal amount of disorder is added in the element values [see Fig.~\ref{fig:S7_Spectra}b for $\pm 1\%$ disorder], characteristic resonances in the spectrum become non-degenerate and slightly split. This case closely resembles experimental results  as seen from the comparison with Fig.~3 in the main text. We call this additional broadening of spectrum {\it inhomogeneous} since it is caused by the fluctuation of element values leading to a change in frequency of particular mode from one node to another, thus rendering a set of splitting peaks instead of a single resonance. Along with this change in the resonance spectrum, a slight breaking of the characteristic symmetry of modes in the circuit is observed, Fig.~\ref{fig:S8_Modes}e-h, since randomly distributed values break the initial $D_{4}$ symmetry of the model. It is worth noting that such fluctuations in values of circuit components lead to larger {\it off-diagonal} disorder in the associated tight-binding model, that is, $2\%$ fluctuations in tunneling amplitudes $J=0.99...1.01$, $K=M=3.96...4.04$, and, moreover, to the emergence of {\it diagonal} $2\%$ disorder in on-site resonance frequencies $f_0$ fluctuating in the range $32.72...33.38~{\rm kHz}$, as follows from Eq.(4) in the main text.

Finally, the spectrum of the strongly disordered circuit with $\pm 10\%$ fluctuations in element values demonstrates considerable splitting of resonances of bulk and edge states, and band gap becomes smaller. It is also seen that the modes become much more asymmetric compared to $\pm 1\%$ disorder and start to hybridize, Fig.~\ref{fig:S8_Modes}i-l. In particular, the hybridization of topological corner state with bulk resonances in observed in Fig.~\ref{fig:S8_Modes}j. Indeed, Fig.~\ref{fig:S7_Spectra}c features bulk curves inside the bandgap, that overlap with the resonance of the corner state and likely correspond to those hybridized resonances occurring in Fig.~\ref{fig:S8_Modes}j.

Besides, $\pm 10\%$ fluctuations in the values of circuit components correspond to the fluctuations of tunneling amplitudes in the ranges $J=0.90...1.10$, $K=M=3.60...4.40$, thus, leading to a $\pm 20\%$ disorder in the initial tight-binding model. Also, the on-site frequencies $f_0$ fluctuate in the range $30.05...36.70$~kHz, corresponding to the presence of strong $\pm 20\%$ diagonal disorder as well. However, corner state remains isolated and still demonstrates the largest Q-factor compared to the other resonances, Fig.~\ref{fig:S7_Spectra}c, illustrating the robustness of higher-order corner states.

\newpage
\section{Supplementary Note 7 -- Retrieval of the topological invariant in experiments}
\label{sec:Circuit_Invariant}

\begin{table}
\begin{center}
\begin{tabular}{c c c c}
\toprule
 & $\Gamma$ & $X$ & $M$\\
\toprule
 $|\bra{U}\ket{\psi_{1}}|$ & 0.9052 & 0.9678 & 0.9996\\
 $|\bra{U}\ket{\psi_{2}}|$ & 0.2567 & 0.1781 & 0.0105\\
 $|\bra{U}\ket{\psi_{3}}|$ & 0.2764 & 0.1190 & 0.0271\\
 $|\bra{U}\ket{\psi_{4}}|$ & 0.1957 & 0.1319 & 0.0074
\end{tabular}
\caption{\label{tab:Overlaps} Projections $|\bra{\psi}\ket{U}|$ of the experimentally reconstructed Bloch function $\ket{U}$ on the theoretical Bloch functions $\psi_{1}$, $\psi_{2}$, $\psi_{3}$, and $\psi_{4}$ at points $\Gamma$ ($k_{x}=0,k_{y}=0$), $X$ ($k_{x}=\pi,k_{y}=0$), and $M$ ($k_{x}=\pi,k_{y}=\pi$), respectively. Theoretically calculated states are sorted in the ascending order, i.e. $\psi_{1}$ corresponds to the lowest-energy bulk state below the bandgap.}
\end{center}
\end{table}

Since the evaluation of topological invariant relies in our case on the study of the Bloch functions (see Supplementary Note 2), we first retrieve these amplitudes from the experiment. To this end, we excite the circuit with the harmonic signal having amplitude $U_{\rm ext}=50~{\rm mV}$ (peak-to-peak voltage $100~{\rm mV}$) in the frequency range $1...20~{\rm kHz}$ applying the signal between node $(5,5)$ and ground. Then, keeping the driving signal at node $(5,5)$, we measure complex voltages between all nodes of the circuit taking node $(5,5)$ as a phase reference. In this way, we obtain a full map of voltages between nodes of the circuit and ground, taking into account their relative phases.

\begin{center}
\begin{figure}
\begin{minipage}{\textwidth}
    \centerline{\includegraphics[width=14cm]{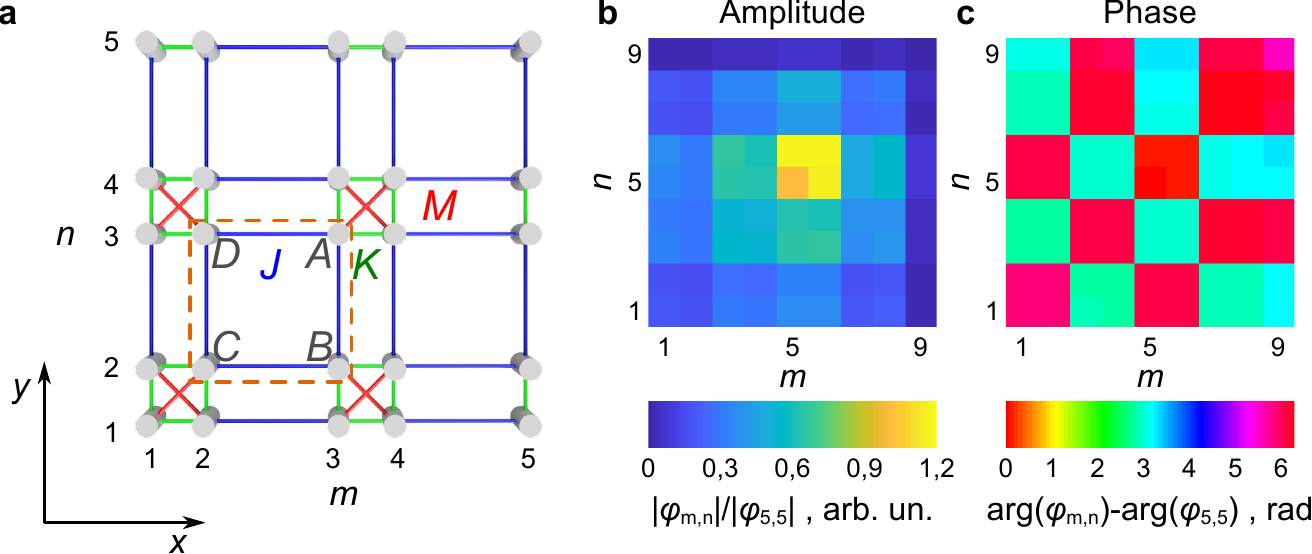}}
    \caption{\textbf{a} Sketch of the extended SSH model with the considered elementary cell with intra-cell couplings $J$ shown with dashed orange line. Labels $A$, $B$, $C$, and $D$ denote the sites of the elementary cell corresponding to four sub-lattices. \textbf{b} Amplitudes of voltages induced between nodes of the circuit and ground when node $(5,5)$ is driven with the harmonic signal with amplitude $50~\rm{mV}$ and frequency $f=18616~{\rm Hz}$. \textbf{c} Phase differences between potentials at nodes of the circuit and the driven node $(5,5)$.\label{fig:S9_Invariant}}
\end{minipage}
\end{figure}
\end{center}

Next, we select a bulk resonance at the frequency $f=18616~{\rm Hz}$ corresponding to the state below bandgap in the extended SSH model. In order to obtain Bloch amplitude of this bulk state, we perform a Fourier transform:
\begin{align*}
    U_{A} &= \sum_{m,n=1}^{N}\varphi_{mn}{\rm e}^{-ik_{x}(m-1)/2 - ik_{y}(n-1)/2}, \quad m,n \in {\rm odd},\\
    U_{B} &= \sum_{m,n=1}^{N}\varphi_{mn}{\rm exp}^{(-ik_{x}(m-1)/2 - ik_{y}(n-2)/2)}, \quad m \in {\rm odd}, n \in {\rm even}\\
    U_{C} &= \sum_{m,n=1}^{N}\varphi_{mn}{\rm exp}^{(-ik_{x}(m-2)/2 - ik_{y}(n-2)/2)}, \quad m,n \in {\rm even},\\
    U_{D} &= \sum_{m,n=1}^{N}\varphi_{mn}{\rm exp}^{(-ik_{x}(m-2)/2 - ik_{y}(n-1)/2)}, \quad m \in {\rm even}, n \in {\rm odd},
\end{align*}
where $\phi_{mn}$ are complex-valued voltages between nodes with coordinates $(m,n)$ and ground, indices $A$, $B$, $C$, and $D$ label four sites of the unit cell with $J$ being intra-cell links, Fig.~\ref{fig:S9_Invariant}a, and wave vector components $k_{x,y}$ are directed along the axes shown in Fig.~\ref{fig:S9_Invariant}a. Applying such a transform to the voltage distribution shown in Fig.~\ref{fig:S9_Invariant}b-c, we obtain a four-component  Bloch amplitude $U(k_{x}, k_{y}) = (U_{A}, U_{B}, U_{C}, U_{D})^{\rm T}$ which depends on Bloch wave vector ${\bf k}$. For high-symmetry points $\Gamma$ ($k_{x}=0, k_{y}=0$), $X$ ($k_{x}=\pi, k_{y}=0$), and $M$ ($k_{x}=\pi, k_{y}=\pi$) we retrieve the following  Bloch amplitudes of the bulk state below the bandgap:

\begin{equation}
    U(0,0) = 
    \begin{pmatrix}
        0.96 + 0.06i \\ 0.39 - 0.61i \\ 1.15 - 0.03i \\ 1.11 + 0.15i
    \end{pmatrix}, \quad
    U(\pi,0) = 
    \begin{pmatrix}
        1.04 + 0.44i \\ -0.63 - 0.02i \\ -0.92 - 0.42i \\ 1.04 + 0.47i
    \end{pmatrix}, \quad
    U(\pi,\pi) = 
    \begin{pmatrix}
        -0.99 + 0.12i \\ 0.94 - 0.15i \\ -1.00 + 0.13i \\ 1.02 - 0.13i
    \end{pmatrix}.
\end{equation}

For the Bloch amplitudes above, we use the same normalization $\sum_{j=1}^{4}|U_{j}|^2=4$  as for the theoretical amplitudes in Supplementary Note 2. Comparing the obtained amplitudes with theoretical Bloch functions for the same bulk state below the bandgap
\begin{equation}
    \psi(0,0) = 
    \begin{pmatrix}
        1\\ 1 \\ 1 \\ 1
    \end{pmatrix}, \quad
    \psi(\pi,0) = 
    \begin{pmatrix}
        1\\ -1 \\ -1 \\ 1
    \end{pmatrix}, \quad
    \psi(\pi,\pi) = 
    \begin{pmatrix}
        -1\\ 1 \\ -1 \\ 1
    \end{pmatrix}.
\end{equation}
one can observe a very good agreement for the $M$-point and a reasonable correspondence for  $\Gamma$- and $X$-points. 

To eliminate the degree of freedom, related to the choice of the phase of the Bloch functions, we project the retrieved Bloch amplitudes on the exact Bloch functions for all four bulk states in points $\Gamma$, $X$, and $M$. The respective products are provided in the Table~\ref{tab:Overlaps} and highlight that the experimentally obtained functions, indeed, correspond to the bulk state $\ket{\psi_1}$ below the bandgap and weakly overlap with the other Bloch functions.


Thus, the entire procedure outlined in Supplementary Note 2 can be repeated for the retrieved approximate Bloch amplitudes, yielding associated eigenvalues of the rotation operator $p(R_{2})=1$ and $p(R_{4})=1$ at $\Gamma$-point, $p(R_{2})=2$ at $X$-point, and $p(R_{4})=3$ at $M$-point, resulting in the topological invariant value
\begin{equation}
    \chi^{(4)}= 
    \begin{pmatrix}
        -1\\ -1 \\ 0
    \end{pmatrix},
\end{equation}
which indicates the presence of a topological higher-order corner state in the experimentally realized electric circuit.